 \documentclass[11pt]{article}

\usepackage{amsmath,amsfonts,amssymb}

 \usepackage[numbers,sort&compress]{natbib}

\usepackage{hyperref}
\hypersetup{
   colorlinks   =  true,
    linkcolor    = blue,
    citecolor    = red,
    urlcolor = blue,
}

\newcommand{\sect}[1]{\setcounter{equation}{0}\section{#1}}
\newcommand{\subsect}[1]{\subsection{#1}}

\renewcommand{\theequation}{\arabic{section}.\arabic{equation}}

\newcommand\be{\begin{equation}}
\newcommand\ee{\end{equation}}
\newcommand\bea{\begin{eqnarray}}
\newcommand\eea{\end{eqnarray}}

  \newcommand\dd{{\rm d}}
  \def\eee{{\rm e}}
 \def\gam{\Theta}
 
  \def\>#1{{\bf #1}} 
\DeclareMathOperator\spn{span}
 \def\mmm{{\cal M}} 
 \def\kk{k} 

   \def\ss{s} 

\begin{document}

 
 \noindent
 {\Large \bf  
Exact solutions and superposition rules for Hamiltonian systems\\[6pt]  generalizing time-dependent SIS epidemic models with   stochastic  \\[6pt] fluctuations} 
 

\medskip

\begin{center}

{\sc  Rutwig Campoamor-Stursberg$^{1,2}$, Eduardo Fern\'andez-Saiz$^{3}$\\[2pt] and Francisco J.~Herranz$^4$}

\end{center}

\medskip

 \noindent
$^1$ Instituto de Matem\'atica Interdisciplinar, Universidad Complutense de Madrid, E-28040 Madrid,  Spain

\noindent 
$^2$ Departamento de Geometr\'{\i}a y Topolog\'{\i}a,  Facultad de Ciencias 
Matem\'aticas, Universidad Complutense de Madrid, Plaza de Ciencias 3, E-28040 Madrid, Spain

\noindent 
{$^3$ Department of Mathematics and Data Science, Universidad San Pablo-CEU,  Campus de Montepr\'incipe,  E-28925 Alcorc\'on, Spain}

\noindent
{$^4$ Departamento de F\'isica, Universidad de Burgos, 
E-09001 Burgos, Spain}

 \medskip
 
\noindent  E-mail: {\small
 \href{mailto:rutwig@ucm.es}{rutwig@ucm.es}, \href{mailto:eduardo.fernandezsaiz@ceu.es}{eduardo.fernandezsaiz@ceu.es}, \href{mailto:fjherranz@ubu.es}{fjherranz@ubu.es} 
}
 
\medskip

\begin{abstract}
\noindent
Using the theory of  Lie--Hamilton systems, formal generalized time-dependent Hamiltonian systems that enlarge a recently proposed SIS epidemic model with a variable infection rate are considered. It is shown that, independently on the particular interpretation of the time-dependent coefficients, these systems generally admit an exact solution, up to the case of the maximal extension within the classification of Lie--Hamilton systems, for which a superposition rule is constructed. The method provides the algebraic frame to which any SIS epidemic model that preserves the above mentioned properties is subjected. In particular, we obtain exact solutions for generalized SIS Hamiltonian models based on the book  and oscillator algebras, denoted by $\mathfrak{b}_2$ and $\mathfrak{h}_4$, respectively. The last generalization corresponds to a SIS system possessing the so-called two-photon algebra symmetry $\mathfrak{h}_6$, according to the embedding chain $\mathfrak{b}_2\subset \mathfrak{h}_4\subset \mathfrak{h}_6$, for which an exact solution cannot generally be found, but a nonlinear superposition rule is explicitly given.
\end{abstract}
\medskip
\medskip

 

\medskip
 
   
 \newpage

\tableofcontents


\sect{Introduction}

  The mathematical theory of epidemics, although formally established as an independent discipline through the pioneering work of  Brownlee, Hamer, Lotka  and Ross at the beginning of the XXth Century, can actually be traced back to the XVIIIth Century, when  Bernoulli 
first proposed the use of differential equations to study the population dynamics of infectious diseases (see e.g.~\cite{Bai,Hethcote} and references therein). Among the various models proposed from the 1930s onwards, combining both analytical and statistical methods, the so-called compartmental models play a relevant role within epidemiological dynamics, with each of the variables or compartments (hence suggesting the terminology) representing a specific stage of contagion. Basic   important examples are given by the well-known SIR models with three compartments.\footnote{Here ``S" stands for the individuals susceptible to the disease, ``I" designates the infected individuals, while ``R" stands for the recovered ones.} The time-evolution of these variables is described in terms of a non-autonomous system of ordinary differential equations \cite{Walt}. There are many variants to  this approach, as studied e.g.~by Kermack and McKendrick \cite{KK} or Reed and Frost \cite{Ab52}, in which additional constraints are introduced, as in the SIRS model \cite{Miller}, in which immunity only lasts for a short period of time, or the MSIR model, 
which takes into account additional assumptions on immunity (see \cite{Bai} and references therein for further details). The simplest assumption are given by the SIS model, in which the individuals in the population do not acquire immunity after infection, implying that they remain susceptible to be infected again \cite{NakamuraMartinez}. Albeit the model appears to be quite simple, the introduction of fluctuations leads to more realistic realizations, although these additional constraints are not entirely obvious. One recurrent tool to take account for fluctuations is to consider stochastic variables, hence to describe the model in terms of stochastic differential equations \cite{Bar,Bun,Or3}. For the particular case of SIS models, the use of the Hamiltonian formalism has led to an improved set of differential equations for the mean and variance of infected individuals \cite{NakamuraMartinez}. In the latter work, the spreading of the disease is treated in terms of Markov chains, in which at most one single recovery or transmission occurs in an infinitesimal interval, and using the mean density of individuals and the variance as dynamical variables. The fluctuations included following this approach correspond to a truncated stochastic expansion, where higher statistical moments are neglected. 
Other alternative approaches to epidemiological models based on the Hamiltonian machinery can be found e.g.~in~\cite{NuLe04,Ba2020,Covid}, where exact solutions of a SIR model have been  obtained in~\cite{Bohner2019,Ba2021}. An approach without a Hamiltonian perspective can be found in~\cite{Kopfova1,Kopfova2}. The Hamiltonian model in \cite{NakamuraMartinez}, in spite of the stochastic nature of the fluctuation, can still be treated as a deterministic system. Genuine stochastic (Hamiltonian) systems, i.e. stochastic differential equations defined on complete probability spaces, have been studied extensively \cite{Or3,Or1,Or2} and successfully applied to SIS models (see \cite{Pan,Otu} and references therein). In this context, the consideration of stochastic parameters/variables allows to apply the formalism to other systems that can be modeled as a contagion-like process, such as information transmission models, networks or one-dimensional diffusion processes \cite{Otu,groh}. This fact legitimates that SIS models can be generalized by introducing time-dependent parameters. 

In \cite{Covid} the authors showed how the machinery of Lie--Hamilton  (LH in short)  systems~\cite{CLS13, BCHLS13Ham, LH2015, BHLS, LuSa} can be applied to the study of SIS models, providing a superposition principle for a time-dependent generalization of the 
model studied in \cite{NakamuraMartinez}. From the perspective of LH systems, further generalizations of the underlying Hamiltonian system that preserve the property that one parameter is identified with a variable infection rate are conceivable, which provide formally extensions of these epidemic models once the additional functions have been identified with relevant epidemiological parameters. In this sense, the time-dependent Hamiltonians considered in this work introduce some semi-stochatic effects by means of fluctuations. It is not our purpose to analyze the epidemiological validity of such extensions, but to present the formal analytical frame to which such models are subjected, study the existence of exact solutions and, alternatively, the existence of nonlinear superposition principles that can be computationally more efficient than a direct attempt of integrating the system.    

We recall that Lie systems~\cite{LuSa,PW,CGM07} are mainly characterized by the existence of (nonlinear) superposition rules, allowing a complete description of the solution of a system of differential  equations in terms of a certain number of particular solutions and significant constants. The superposition principles are themselves deeply related to the existence of a finite-dimensional Lie algebra of vector fields associated to the system, which further leads to a realization of the system in terms of a sum of the Lie algebra generators with time-dependent coefficients \cite{LuSa}. As shown in \cite{Or4}, using the integrability properties of distributions generated by vector fields, the formalism can be extended to the case of stochastic differential equations, leading to a stochastic version of the Lie--Scheffers theorem and corresponding (local) superposition rules.

A particularly interesting case arises whenever the vector fields spanning the Lie algebra are Hamiltonian vector fields with respect to a symplectic structure, a feature that allows us to treat the system as a classical Hamiltonian system \cite{ARN}. Systems of this type, called LH systems, have been extensively studied in low dimensions (see~\cite{CLS13, BCHLS13Ham, LH2015, BHLS, LuSa} and references therein) and constitute a powerful technique for the analysis of differential equations. One of the advantages of the compatibility with a symplectic structure resides in the fact that the so-called coalgebra formalism of (super-)integrable systems~\cite{BBHMR09} can be applied to systematize the construction of time-independent constants of the motion~\cite{BCHLS13Ham}, which finally provides a superposition principle through algebraic computations, so avoiding the cumbersome integration of systems of PDEs or ODEs.

As mentioned, the main objective of this work is to embed the SIS Hamiltonian models of~\cite{NakamuraMartinez}  and~\cite{Covid} into larger Hamiltonian systems, such that the underlying symplectic structure is preserved, and determine to which extent exact solutions of these formal generalizations can be obtained, alternatively applying the techniques of LH systems to derive suitable superposition rules. The potential interpretation of the various new  time-dependent  coefficients introduced in this way is not considered, nor the applicability to systems beyond epidemiological models, such as the spread of information over networks or numerical methods. In Section~\ref{s2} we briefly review the SIS Hamiltonian model proposed by Nakamura and Martinez in~\cite{NakamuraMartinez}, which possesses a constant infection rate. In Section~\ref{s3}, we reconsider the generalization proposed in~\cite{Covid}  and based on the introduction of a variable infection rate. The striking point is that the $t$-dependent Hamiltonian studied in~\cite{Covid} gives rise to a natural generalization by adding a second arbitrary  $t$-dependent parameter in such a manner that the resulting Hamiltonian can be  related to the so called book LH algebra $\mathfrak{b}_2$~\cite{BHLS,Ballesteros6}. The classification of LH systems on the plane $\mathbb R^2$~\cite{LH2015,BHLS} allows us to obtain a canonical transformation between the generic Cartesian coordinates $(x,y)$ and the mean  density of infected individuals $\langle\rho\rangle$ and the variance $\sigma^2$ of this generalized Hamiltonian model. Using the geometric and algebraic properties of LH systems we derive an exact explicit solution for this book Hamiltonian, which holds in particular for the model proposed in \cite{Covid}, where no exact solution was given. 

Using the subalgebra lattice of the classification of planar LH systems~\cite{LH2015,BHLS}, a second extension of the book LH algebra $\mathfrak{b}_2$ to the oscillator algebra $\mathfrak{b}_2 \subset\mathfrak{h}_4$ is considered introducing an additional $t$-dependent parameter. This case is considered in Section~\ref{s4} in full detail. In particular, we present an exact solution for this new oscillator Hamiltonian, providing alternatively a superposition rule that, albeit not required because of the exact solution, will be convenient in the context of extensions of the system to higher dimensional LH systems. In Section~\ref{s5} we consider the maximal embedding chain (within the LH classification) $\mathfrak{b}_2 \subset\mathfrak{h}_4 \subset\mathfrak{h}_6$ that leads to the so-called two-photon Lie algebra  $\mathfrak{h}_6$~\cite{Gilmore,BBF09}. Although in this case the first-order system of differential equations is still linear (in the Cartesian coordinates) with variable coefficients, a direct integration will be generally no more possible or computationally feasible, due to its equivalence with a reduced system of Riccati type. This justifies the construction of a superposition principle, eventually allowing a description of the solution in terms of three particular solutions. 
 
Finally, some open problems and future prospects are discussed in Section~\ref{s6}. For completeness in the exposition, in the Appendix we briefly recall the fundamental properties of LH systems, the technical details of which can be found in~\cite{CLS13,BCHLS13Ham,LuSa,BHLS,LH2015}.


 \sect{A  SIS  Hamiltonian model with stochastic fluctuations}
 \label{s2}

  As mentioned previously, one of the main premises of the SIS model is that all individuals are still susceptible to the infection after recovery, which implies that they do not acquire immunity. From this condition we conclude that the model can be described only using two compartmental variables. The first one, $I$, corresponds to the number of infected individuals, whereas the second compartment $S$ describes the number of individuals susceptible to the infection at a given time.
The model assumes a large population of size $N$, a random mixture of the population, a regular and stationary age distribution, as well as one single contaminating agent. The chances of infection for any individual extend through the whole course of the epidemic. In this context, one relevant variable is given by the density $\rho=\rho(\tau)$ of infected individuals, depending on the time parameter $\tau$ that takes values in the interval $[0,1]$. The density of infected individuals decreases according to a rate $\gamma \rho$, with $\gamma$ denoting the recovery rate, while the growth rate of new infections is proportional to $\alpha \rho(1-\rho)$, with a transmission rate parameter $\alpha$. The equation that describes the variation of infected individuals taking into account both contribution factors is given by:
\begin{equation}\label{compsis1}
\frac{\dd \rho}{\dd\tau}=\alpha \rho(1-\rho)-\gamma \rho.
\end{equation}
One can redefine the timescale as $t := \alpha \tau$ (where $\alpha\neq 0$) and introduce the constant $\rho_0:=1-\gamma/\alpha$, so that we can reformulate equation  (\ref{compsis1}) as
\begin{equation}\label{sismodel0}
\frac{\dd\rho}{\dd t}= \rho(\rho_0-\rho).
\end{equation}
Clearly, the equilibrium density is reached if either $\rho=0$ or $\rho=\rho_0$. 
 
  One major drawback of the model is that it assumes that each individual of the 
population lives to a certain maximal age $L$, and that for each age $a<L$, the number of individuals of age $a$ is the same. While such a  homogeneity assumption seems feasible in developed countries, where infant mortality is quite low, the hypothesis cannot be assumed realistically to hold for developing countries. To circumvent this difficulty, 
it is reasonable to introduce probability functions at some point, in order to allow a random variation over time in one or more of the inputs. Some experimental results show conclusively that temporal fluctuations can drastically modify
the prevalence of pathogens and the spatial heterogeneity (see e.g.~\cite{real,duncan}).   

 In order to introduce fluctuations in the SIS model, we follow the general ansatz proposed by Nakamura and Martinez  in~\cite{NakamuraMartinez}, where the spreading of the disease is interpreted as a Markov chain in discrete time with at most one single recovery or transmission occurring in each infinitesimal interval. Therefore, the equations for the instantaneous mean density of infected individuals $\langle\rho\rangle$ and the variance $\sigma^2=\langle\rho^2\rangle-\langle\rho\rangle^2$ are \cite{kiss}:
\begin{equation}\label{sismodel1}
\begin{split}
\frac{\dd\langle\rho\rangle}{\dd t}&=\langle\rho\rangle \left( \rho_0-\langle\rho\rangle \right)-\sigma^2 ,\\
\frac{\dd\sigma^2}{\dd t}&=2\sigma^2 \left(\rho_0+\langle\rho\rangle\right)-2 \Delta_3 +\frac{1}{N}\langle\rho (1-\rho)\rangle+\frac{1}{N }\, (1-\rho_0)\langle\rho\rangle,
\end{split}
\end{equation}
where $\Delta_3 =\langle\rho^3\rangle-\langle\rho\rangle^3$ and the variance in the first equation slows down the growth rate of $\langle\rho\rangle$, recalling the Allee effect \cite{Ribe}. Equations \eqref{sismodel0} and \eqref{sismodel1} are equivalent whenever $\sigma$ becomes negligible when compared to $\langle\rho\rangle$, and  ignoring higher statistical moments, so that the dynamical system describes a Gaussian variable evolving along time. This implies that $\Delta_3 \simeq3\sigma^2\langle\rho \rangle$, thus for a sufficiently large number of individuals $N$, the resulting equations read
\begin{equation}\label{sislabellog}
\begin{split}
\frac{\dd \ln{\langle\rho\rangle}}{\dd t}&=\rho_0-\langle\rho\rangle-\frac{\sigma^2}{\langle\rho\rangle},\\
\frac{1}{2}\frac{\dd \ln{\sigma^2}}{\dd t}&=\rho_0-2\langle\rho\rangle.
\end{split}
\end{equation}
With the introduced fluctuations, the equations \eqref{sislabellog} correspond with a stochastic expansion, according to \cite{vilar}. In this situation, we assume that the density $\rho=\langle \rho\rangle +\eta$ is properly described by the instantaneous average as well as some noise function $\eta$. For consistency, it is further assumed that $\langle \eta\rangle = 0$ and $\langle \eta^2\rangle = \sigma^2$. The latter system \eqref{sislabellog} allows a Hamiltonian formulation, with the phase space variables given by the mean  density of infected individuals $\langle\rho\rangle$ and the variance $\sigma^2$. Considering 
  \be
  q=\langle\rho\rangle, \qquad p=\sigma^{-1}
  \label{z1}
  \ee
   as  dynamical variables~\cite{NakamuraMartinez}, the system \eqref{sislabellog} adopts the form 
\begin{equation}\label{sismodel3}
\begin{split}
\frac{\dd q}{\dd t}&=\rho_0q-q^2-\frac{1}{p^2},
\\
\frac{\dd p}{\dd t}&=-\rho_0p+2qp.
\end{split}
\end{equation}
It is straightforward to verify that these are the canonical equations associated to the Hamiltonian
\begin{equation}\label{hamiltoniansis}
H=qp\left(\rho_0-q\right)+\frac{1}{p}.
\end{equation}
As shown in~\cite{NakamuraMartinez}, the system (\ref{sislabellog}) can be solved exactly, providing the solution 
\begin{equation}
\begin{split}
\langle\rho(t)\rangle&=\frac{ \rho_0\bigl(1+ \tilde c_1 \eee^{-\rho_0 t} \bigr)}{1+ 2 \tilde c_1  \eee^{-\rho_0 t} + \tilde c_2  \eee^{-2\rho_0 t} } \, ,
\\[4pt]
\sigma^{2}(t)&= \frac{\langle\rho(t)\rangle^2 \bigl(\tilde c_1^2 - \tilde c_2  \bigr)    \eee^{-2\rho_0 t}   }{\bigl(1+ \tilde c_1 \eee^{-\rho_0 t} \bigr)^2 } \, ,
\end{split}
\label{solution0}
\end{equation}
where $\tilde c_1$ and $\tilde c_2$ are two   constants depending on the initial conditions.  In terms of the canonical variables $(q,p)$  in   (\ref{z1}), the solution of the system (\ref{sismodel3}) adopts the following expression
\begin{equation}
\begin{split}
q(t)&=\frac{ \rho_0\bigl(1+ \tilde c_1 \eee^{-\rho_0 t} \bigr)}{1+ 2 \tilde c_1  \eee^{-\rho_0 t} + \tilde c_2  \eee^{-2\rho_0 t} } \, ,
\\[4pt]
p(t)&= \frac{ 1+ 2 \tilde c_1  \eee^{-\rho_0 t} + \tilde c_2  \eee^{-2\rho_0 t}  }{\rho_0 \sqrt{\tilde c_1^2- \tilde c_2}\,  \eee^{-\rho_0 t} }\, .
\end{split}
\label{solution1}
\end{equation}
 For a detailed analysis of all of the above results we refer to~\cite{NakamuraMartinez}.


\sect{Generalization of the     SIS  Hamiltonian model with a variable\\ infection rate}
 \label{s3}

Quite recently, a generalization of the Hamiltonian (\ref{hamiltoniansis}) was proposed in~\cite{Covid} by considering a $t$-dependent      infection rate through a smooth function $\rho_0(t)$, which amounts to introduce a $t$-dependent basic reproduction number $R_0(t)$ that is actually observed in more accurate epidemic models~\cite{Cui,Gao,Driessche}. The remarkable point, as  was explicitly shown in~\cite{Covid}, is that the resulting $t$-dependent Hamiltonian inherits the structure of a LH system~\cite{CLS13, BCHLS13Ham, LH2015, BHLS, LuSa}. In the following we summarize the main results.

If we consider that $\rho_0=\rho_0(t)$ and insert it into equations (\ref{sismodel3}), we obtain the system
\begin{equation}\label{sismodel3b}
\begin{split}
\frac{\dd q}{\dd t}&=\rho_0(t)\, q-q^2-\frac{1}{p^2} ,
\\
\frac{\dd p}{\dd t}&=-\rho_0(t)\, p+2qp.
\end{split}
\end{equation}
The dynamics of the resulting system is easily described in terms of a $t$-dependent vector field  (see (\ref{TDEPA}) in the Appendix)
\begin{equation}\label{sisliesystem}
{\bf X}=\rho_0(t){\bf X}_A+{\bf X}_B, \qquad {\bf X}_A=q\,\frac{\partial}{\partial q}-p\,\frac{\partial}{\partial p},\qquad {\bf X}_B=-\left(q^2+\frac{1}{p^2}\right)\frac{\partial}{\partial q}+2 q p\,\frac{\partial}{\partial p},
\end{equation}
and such that the commutation rule
\be
[{\bf X}_A,{\bf X}_B]={\bf X}_B 
\label{sa}
\ee
holds. This implies that the generalized SIS model (\ref{sismodel3b}) determines a Lie system~\cite{LuSa,LSc,PW}, the  Vessiot--Guldberg algebra of which is isomorphic to the so-called book algebra, here denoted by $\mathfrak{b}_2$. In this context, ${\bf X}_A$ can be regarded as a dilation generator, while ${\bf X}_B$ corresponds to a translation generator. 

In addition, as $(q,p)$ are canonical variables, we  can consider the usual symplectic form in  the  phase space 
\be
\omega = \dd q \wedge\dd p,
\label{sb}
\ee
and  compute  the corresponding  Hamiltonian functions  $h_A,\,h_B $ associated to the vector fields ${\bf X}_A,\,{\bf X}_B$  through the contraction or inner product of $\omega$, that is,
\be
\iota_{{\bf X}_i}\omega={\rm d}h_i ,\qquad  i=A, B.
\label{sb2}
\ee
The   functions  $h_A,\,h_B $ are given by
\begin{equation}\label{hamfunctsis}
h_A=qp,\qquad h_B= \frac{1-q^2p^2}{p},
\end{equation}
and satisfy the Poisson bracket 
 \be
 \{h_A,h_B\}_\omega=-h_B .
 \label{sd}
 \ee
   Thus ${\bf X}_A,\, {\bf X}_B$ are Hamiltonian vector fields and the $t$-dependent  Hamiltonian of the  generalized SIS model   (\ref{sismodel3b}) reads  (see (\ref{ag}))
\begin{equation}\label{hamcovid}
h=\rho_0(t)h_A+h_B=\rho_0(t)qp + \frac{1-q^2p^2}{p},
\end{equation}
reproducing the Hamiltonian function (\ref{hamiltoniansis}) proposed in \cite{NakamuraMartinez} for a constant $\rho_0$. This  shows that both cases can be studied by means of a unified geometrical approach. In consequence, the 
 generalized SIS model   (\ref{sismodel3b})  is not merely a Lie system, but also an LH one~\cite{CLS13, BCHLS13Ham, LH2015, BHLS, LuSa}, as was already proved in~\cite{Covid}.


\subsect{Generalized    SIS Hamiltonian   from the    book algebra}
 \label{s31}

By taking into account the LH theory,  the SIS Hamiltonian (\ref{hamcovid}) can be further generalized through the introduction of a second arbitrary $t$-dependent parameter $b(t)$ in the form
\begin{equation}
h=\rho_0(t)h_A+b(t) h_B=\rho_0(t)qp+b(t) \left( \frac{1-q^2p^2}{p}\right) ,
 \label{se}
\end{equation}
leading to the following system of differential equations
\begin{equation} 
\begin{split}
\frac{\dd q}{\dd t}&=\rho_0(t) q - b(t) \left(  q^2+\frac{1}{p^2} \right)  ,
\\
\frac{\dd p}{\dd t}&=-\rho_0(t) p+2b(t)  qp.
\end{split}
 \label{sf}
\end{equation}
Clearly, the associated  $t$-dependent vector field is just 
\be
{\bf X}=\rho_0(t){\bf X}_A+b(t){\bf X}_B,
 \label{sg}
\ee
 with 
$ {\bf X}_A $ and $ {\bf X}_B$ given in (\ref{sisliesystem}).

Summing up, for any choice of the parameters $\rho_0(t)$ and $ b(t)$ the  differential equations (\ref{sf})   always determine  an LH system  with  an associated    $\mathfrak{b}_2$-LH algebra.  For this reason, we can say that  (\ref{sf}) 
is a generalized (time-dependent) Hamiltonian from the book LH algebra that includes the SIS epidemic model (\ref{sismodel3b}) studied in~\cite{Covid} for the values $b(t)\equiv 1$, as well as the model (\ref{sismodel3}) introduced in~\cite{NakamuraMartinez} for the values $\rho_0(t)\equiv \rho_0$ and $b(t)\equiv 1$.

We recall that  the   $\mathfrak{b}_2$-LH algebra  appears within the classification of planar   LH systems presented in~\cite{LH2015,BHLS} as the class I$_{14A}^{r=1}\simeq \mathbb{R} \ltimes \mathbb{R}\simeq \mathfrak{b}_2$. Let us consider the Hamiltonian vector fields  spanning  the $\mathfrak{b}_2$-LH algebra  in terms of the usual Cartesian coordinates $(x,y)\in \mathbb R^2$, as given in~\cite{Ballesteros6}:
 \be
 h_A=xy ,\qquad h_B= -x.
  \label{sh}
 \ee
These verify the Poisson bracket (\ref{sd}) with respect to the    standard symplectic form  \be
\omega = \dd x \wedge\dd y .
\label{si}
\ee
The  vector fields associated with (\ref{sh}) are obtained using (\ref{sb2}) and equal 
  \be
 \>X_A=x\, \frac{\partial}{\partial x}- y\, \frac{\partial}{\partial y},  \qquad    \>X_B=\frac{\partial}{\partial y}, 
      \label{si2}    
\ee
satisfying the commutator relation (\ref{sa}). From the classification of LH systems we deduce the existence of a canonical transformation between the variables $(q,p)$ in (\ref{sisliesystem}) and (\ref{hamfunctsis}) and $(x,y)$ in (\ref{sh}) and (\ref{si2}) respectively; this turns out to be
\be
\begin{split}
x&=\frac {q^2 p^2-1} p ,\qquad y= \frac{qp^2}{q^2p^2-1},\\
 q&=\frac{x^2 y}{x^2 y^2 -1} , \qquad  p = \frac{x^2  y^2 -1}x,
\end{split}
\label{sk}
\ee
preserving  the symplectic form 
\be
\omega = \dd x \wedge\dd y= \dd q \wedge\dd p    .
\label{sk2}
\ee
It is worthy to be mentioned that the canonical transformation (\ref{sk}) is quite different from the change of variables considered in~\cite{Covid} (cf.~equation~(34)), as the latter does not preserve the symplectic form. 
Observe that, in absence of the canonical character of such transformation, the variables $q$ and $p$ used   in~\cite{Covid} do not correspond to canonical dynamical variables, hence  their direct interpretation as  the mean  density of infected individuals $\langle\rho\rangle$ and the variance $\sigma^2$ (\ref{z1}) is no more entirely obvious.

  
   \subsect{Exact solution for the book SIS Hamiltonian}
 \label{s32}

The remarkable fact is that the system (\ref{sf}) becomes separable in the coordinates $(x,y)$, as can be routinely verified:
   \begin{equation} 
\frac{\dd x}{\dd t}=\rho_0(t) x ,
\qquad
\frac{\dd y}{\dd t}=-\rho_0(t) y+ b(t)  .
 \label{sl}
\end{equation}
The equations are uncoupled and can be solved in straightforward way by quadrature. Considering again the canonical transformation (\ref{sk}), the corresponding solution of the initial system  (\ref{sf}) is obtained.
   
The explicit solution of the linear system (\ref{sl}) is given by
   \be
\begin{split}
x(t)&=c_1 \,\eee^{\gam(t)},\qquad \gam(t):= \int_a^t\rho_0(s)\dd s ,\\[4pt]
y(t)&=\left( c_2 +  \int_a^t  \eee^{\gam(u )} b(u) \dd u\right)  \eee^{-\gam(t)},
\end{split}
\label{sm}
\ee
where $c_1$ and $c_2$ are the two  constants of integration, determined by the  initial conditions, while $a$ is a real number that ensures the existence of the integrals over the interval $[a,t]$. By introducing the transformation (\ref{sk}) we arrive at the general solution of   (\ref{sf}) for arbitrary $t$-dependent parameters  $\rho_0(t)$ and $b(t)$:
    \be
\begin{split}
q(t)&=\frac{\left( c_2 +  \int_a^t  \eee^{\gam(u )} b(u) \dd u\right)\eee^{\gam(t)} }{\left( c_2 +  \int_a^t  \eee^{\gam(u )} b(u) \dd u\right)^2-c_1^{-2} } \,  ,\\[4pt] 
p(t)&=\left(  c_1\biggl( c_2 +  \int_a^t  \eee^{\gam(u )} b(u) \dd u\biggr)^2- c_1^{-1}  \right)  \eee^{-\gam(t)} \, .
\end{split}
\label{sn}
\ee
Recall that the resulting mean  density of infected individuals and the variance of the model are just $\langle\rho(t)\rangle = q(t)$  and $\sigma^2(t)=1/p^2(t)$  (see  (\ref{z1})).  Observe also that, as a byproduct of the general result above, the exact solution of the system (\ref{sismodel3b}) considered in~\cite{Covid} is easily obtained setting $b(t)\equiv 1$ and keeping a variable $\rho_0(t)$, a fact that complements the results of that work.

Finally,   we fix  $b(t)\equiv 1$ and consider a constant  $\rho_0(t)\equiv \rho_0$. Then we can choose, for instance,  $a=0$ in the integral $\gam(t)$ (\ref{sm}), that is,   $\gam(t)= \rho_0 t$. Hence the solution (\ref{sn}) reduces to
    \be
q(t)=\frac{\rho_0 \left(   \eee^{\rho_0 t}+ c_2\rho_0-1\right)\eee^{\rho_0 t} }{ \left(   \eee^{\rho_0 t}+ c_2\rho_0-1\right)^2-\rho_0^2c_1^{-2} },\qquad 
p(t)=\left(     \frac{ c_1\left(   \eee^{\rho_0 t}+ c_2\rho_0-1\right)^2  } {\rho_0^2}  - \frac 1{c_1}   \right)  \eee^{-\rho_0 t} \, ,
\label{so}
\ee
recovering, as expected, the solution (\ref{solution1}) of the system (\ref{sismodel3})  studied in~\cite{NakamuraMartinez}, 
provided that the constants of integration are redefined as
\be
c_1=\frac{\rho_0}{\sqrt{\tilde c_1^2-\tilde c_2}},\qquad c_2=\frac{\tilde c_1+1}{\rho_0}.
\label{sp}
\ee


\sect{Generalized  Hamiltonian from the oscillator algebra}
 \label{s4}

The classification of   LH systems~\cite{LH2015,BHLS} shows that the two-dimensional book LH algebra appears as a Lie subalgebra of other classes. This allows us to further extend the results of the previous section by considering higher dimensional LH algebras that entail the introduction of additional $t$-dependent parameters $b_i(t)$, and such that the book SIS system (\ref{sf}) is recovered for the vanishing values of these parameters. Among the various LH algebras into which the book algebra can be embedded, the natural candidate is the class I$_{8}$. As a Lie system, this only requires the addition to $X_A$ and $X_B$ in (\ref{si2}) of a new vector field. Keeping  the same notations of~\cite{LH2015,BHLS}, I$_{8}$ can be expressed in terms of the Cartesian coordinates $(x,y)$, such that the the three vector fields are given by
  \be
 \>X_1=\frac{\partial}{\partial x} ,\qquad  \>X_2\equiv    \>X_B=\frac{\partial}{\partial y},\qquad  \>X_3\equiv \>X_A=x\, \frac{\partial}{\partial x}- y\, \frac{\partial}{\partial y} , 
      \label{ta}    
\ee
obeying the commutation rules
\be
[{\bf X}_3,{\bf X}_1]=-{\bf X}_1 , \qquad [{\bf X}_3,{\bf X}_2]={\bf X}_2 , \qquad [{\bf X}_1,{\bf X}_2]=0 , \qquad 
\label{tb}
\ee
thus closing on a Lie algebra  isomorphic to the (1+1)-dimensional Poincar\'e
algebra $\mathfrak{iso}(1, 1)$.  The Lie system $\>X$   determined by the $t$-dependent vector field
\be
{\bf X} =  b_1(t) {\bf X}_1 +  b_2(t) {\bf X}_2+ \rho_0(t) {\bf X}_3 ,
\label{tc} 
\ee
 leads to   the  following (linear) system of differential equations
    \be
\begin{split}
\frac{\dd x}{\dd t}&= \rho_0(t) x+b_1(t)  , \\[2pt]
 \frac{\dd y}{\dd t}&=- \rho_0(t)   y + b_2(t) .
 \end{split}
\label{td}
\ee

The generators $\>X_i$  (\ref{ta}) become Hamiltonian vector fields $h_i$  with respect to the   symplectic form (\ref{si}), namely
\be
h_1=y,\qquad h_2\equiv h_B=-x,\qquad h_3\equiv h_A= x y,\qquad h_0 =1.
\label{te}
\ee
Observe that, in order to ensure that these functions close with respect to the Lie--Poisson bracket, it is necessary to add the central generator $h_0$:
\be
\{h_3,h_1\}_\omega= h_1,\qquad   \{h_3,h_2\}_\omega=- h_2,\qquad  \{h_1,h_2\}_\omega=h_0,\qquad   \{h_0,\cdot\}_\omega=0 .
\label{tf}
\ee
It follows that the resulting LH algebra is isomorphic to the centrally extended Poincar\'e algebra $\overline{\mathfrak{iso}}(1,1)$,  which is also isomorphic  to the   oscillator  algebra $\mathfrak{h}_4$. Within this identification, $h_3$ can be interpreted as the number operator, while $h_1$, $h_2$ can be seen as ladder operators. Alternatively, the differential equations (\ref{td}) can also be obtained through the Hamilton equations with canonical variables $(x,y)$ from the Hamiltonian given by
   \begin{equation}
h= b_1(t)h_1 +  b_2(t) h_2+ \rho_0(t)h_3     .
 \label{tg}
\end{equation}

The canonical transformation (\ref{sk}) allows us to express the vector fields (\ref{ta})  and the  Hamiltonian vector fields (\ref{te})  in  terms of the canonical variables $(q,p)$ in (\ref{z1}):
    \be
\begin{split}
 \>X_1&= - \frac{2qp}{(q^2p^2-1)^2} \,\frac{\partial}{\partial q}+\frac{p^2(q^2p^2+1)}{(q^2p^2-1)^2}\,\frac{\partial}{\partial p}, \qquad \>X_2\equiv   {\bf X}_B=-\left(q^2+\frac{1}{p^2}\right)\frac{\partial}{\partial q}+2 q p\,\frac{\partial}{\partial p},    \\[2pt]
    \>X_3&\equiv  {\bf X}_A=q\,\frac{\partial}{\partial q}-p\,\frac{\partial}{\partial p},
   \\[2pt]
h_1&=\frac{qp^2}{q^2 p^2 -1},\qquad h_2\equiv h_B=\frac{1-q^2p^2}{p},\qquad h_3\equiv h_A= qp,\qquad h_0 =1.
 \end{split}
\label{th}
\ee
This procedure leads to a genuinely coupled nonlinear first-order system of differential equations that generalizes the book SIS system (\ref{sf}) with an additional parameter $b_1(t)$ 
\begin{equation} 
\begin{split}
\frac{\dd q}{\dd t}&=\rho_0(t) q-2b_1(t) \, \frac{qp}{(q^2p^2-1)^2}  - b_2(t) \left(  q^2+\frac{1}{p^2} \right)  ,
\\[2pt]
\frac{\dd p}{\dd t}&=-\rho_0(t) p+b_1(t) \, \frac{p^2(q^2p^2+1)}{(q^2p^2-1)^2}  +2b_2(t)  qp.
\end{split}
 \label{ti}
\end{equation}
This system can be regarded as a generalized time-dependent Hamiltonian from the oscillator LH algebra that extends the SIS epidemic model (\ref{sf}) for the embedding chain $\mathfrak b_2\subset \mathfrak h_4$, or alternatively, a formal Hamiltonian system into which the SIS epidemic model has been embedded.

  
   \subsect{Exact solution for the oscillator time-dependent Hamiltonian}
 \label{s41}

In spite of the apparently complicated structure of the system of oscillatory type (\ref{ti}), an exact solution can still be found taking into account the previous cases and the fact that, by means of the canonical transformation (\ref{sk}), the system decouples in the new variables, but still preserving the symplectic form $\omega$ and hence, the underlying geometric structure. The system (\ref{td}), being linear, can be easily integrated and gives 
   \be
\begin{split}
x(t)&=\left( c_1 +  \int_a^t  \eee^{-\gam(u )} b_1(u) \dd u\right)  \eee^{\gam(t)},\qquad \gam(t):= \int_a^t\rho_0(s)\dd s ,\\[4pt]
y(t)&=\left( c_2 +  \int_a^t  \eee^{\gam(u )} b_2(u) \dd u\right)  \eee^{-\gam(t)},
\end{split}
\label{tj}
\ee
where again $c_1$ and $c_2$ are the two  constants of integration determined by the  initial conditions. By applying the canonical transformation (\ref{sk}) we  obtain the exact solution of (\ref{ti}), that adopts the form
\be
\begin{split}
q(t)&=\frac{ \left( c_1 +  \int_a^t  \eee^{-\gam(u )} b_1(u) \dd u\right) ^2\left( c_2 +  \int_a^t  \eee^{\gam(u )} b_2(u) \dd u\right)\eee^{\gam(t)} }{ \left( c_1 +  \int_a^t  \eee^{-\gam(u )} b_1(u) \dd u\right) ^2\left( c_2 +  \int_a^t  \eee^{\gam(u )} b_2(u) \dd u\right)^2-1 } \,  ,\\[4pt] 
p(t)&=\frac{    \left( \left( c_1 +  \int_a^t  \eee^{-\gam(u )} b_1(u) \dd u\right) ^2\left( c_2 +  \int_a^t  \eee^{\gam(u )} b_2(u) \dd u\right)^2-1 \right)        \eee^{-\gam(t)} }{\left( c_1 +  \int_a^t  \eee^{-\gam(u )} b_1(u) \dd u\right) }    \, .
\end{split}
\label{tk}
\ee
It can be trivially verified that this solution reduces to solution (\ref{sn})  of the book SIS system for $b_1(t)\equiv 0$ and $b_2(t)\equiv b(t)$.

It should be mentioned that the integrability by quadratures of the systems  (\ref{sf}) and (\ref{ti}) can actually be inferred from the structural properties of the book and oscillator algebras as   Vessiot--Guldberg algebras of a Lie system. As shown in \cite{Car1} (see also \cite{Car2,Car3,Car4} and references therein), if the Vessiot--Guldberg algebra is solvable, then the integrability by quadratures is guaranteed.    

  
   \subsect{Superposition rule for the oscillator time-dependent Hamiltonian}
 \label{s42}

As both the generalized book and oscillator systems determined by the  differential equations (\ref{sf}) and (\ref{ti}) are classical Lie systems, they always admit a (nonlinear) superposition rule \cite{LSc} (see (\ref{LS01}) in the Appendix). As these equations also determine LH systems, an explicit superposition rule can be found in terms of $t$-independent constants of the motion constructed by means of the coalgebra formalism presented in~\cite{BCHLS13Ham} (see also~\cite{BHLS,Ballesteros6}), and based on the Casimir operators (see the Appendix). This formalism solely fails for the book algebra $\mathfrak{b}_2$, as this Lie algebra does not admit nonconstant Casimir invariants. However, using appropriate embeddings of the book algebra, such as $\mathfrak{b}_2\subset  \mathfrak{h}_4$, exact solutions (\ref{sn}) and (\ref{tk}) for both the book and oscillator SIS systems can be obtained. According to the general features of LH systems, for the oscillator Hamiltonian (\ref{tg}) we can find a superposition rule, which, particularized to the book SIS Hamiltonian (\ref{se}), would provide a superposition rule for the latter. Albeit deriving a superposition principle for the system (\ref{ti}) is redundant, as the system can be solved explicitly, it is advantageous to compute it for comparison purposes. As will be shown in the next section, the oscillator time-dependent $\mathfrak{h}_4$-Hamiltonian (\ref{tg}) appears itself as a special case of another and last extension, that is,  the two-photon $\mathfrak{h}_6$-system  for which, although an explicit solution of the associated system is formally conceivable, its computational implementation is probably too cumbersome due to a nontrivial coupling. 
 
We now proceed to compute $t$-independent constants of the motion for the oscillator SIS system (\ref{ti}) and deduce a superposition rule.
  Let us consider the oscillator LH algebra expressed in a basis $\{ v_1,v_2,v_3,v_0 \}$ that formally satisfies the same Poisson brackets (\ref{tf}):
  \be
\{v_3,v_1\}= v_1,\qquad   \{v_3,v_2\}=- v_2,\qquad  \{v_1,v_2\}=v_0,\qquad   \{v_0,\cdot\}=0 .
\label{tl}
\ee
There exists a non-trivial quadratic Casimir element given by
\be
C=v_{1}v_{2}+ v_{3}v_{0} ,\qquad \{C ,\cdot\}=0 .
  \label{tm}
\ee
 In order to derive a superposition rule for this case, we require $m=2$ particular solutions and $n=2$ significant constants (see (\ref{LS01})), and hence the indices $k=2,3$ in the application of the coalgebra formalism~\cite{BHLS} (see (\ref{aj})). From (\ref{te}) we construct the Hamiltonian vector fields  $h^{(\kk)}_i$ in Cartesian coordinates $(x,y)\in \mathbb{R}^2$:
 \begin{equation}
\begin{split}
 h^{(1)}_1 &= y_1,\qquad    h^{(1)}_2  =-x_1,\qquad   h^{(1)}_3  =x_1 y_1,\qquad  h^{(1)}_0  =1 ,    \\[2pt]
  h^{(2)}_1 &= y_1+y_2 ,\qquad    h^{(2)}_2  =-x_1-x_2,\qquad   h^{(2)}_3  =x_1 y_1+x_2 y_2,\qquad  h^{(2)}_0  =2 ,    \\[2pt]
  h^{(3)}_1 &= y_1+y_2+y_3 ,\qquad    h^{(3)}_2  =-x_1-x_2-x_3,\qquad   h^{(3)}_3  =x_1 y_1+x_2 y_2+x_3 y_3,\qquad  h^{(3)}_0  =3
 .\end{split}
 \label{tn}
\end{equation}
Each set  satisfies the Poisson brackets (\ref{tf}) with respect to the symplectic form
   \be
\omega = \dd x_1\wedge \dd y_1+\dd x_2\wedge \dd y_2+\dd x_3\wedge \dd y_3 .
\label{to}
\ee
From this result, using the Casimir invariant (\ref{tm}), we obtain two  $t$-independent constants of the motion $F^{(k)}$ (\ref{ak}) for the diagonal prolongation $\widetilde{ \>X}^3$   to $(\mathbb{R}^2)^3$~\cite{BHLS,Ballesteros6}:
 \begin{equation}
\begin{split}
F^{(2)}&= (x_{1}-x_{2})   (y_{1}-y_{2}) ,  \\[2pt]
F^{(3)}&= (x_{1}-x_{2})   (y_{1}-y_{2})+(x_{1}-x_{3})   (y_{1}-y_{3})+(x_{2}-x_{3})   (y_{2}-y_{3}).  
\end{split}
 \label{tp}
\end{equation}
$F^{(2)}$ and $F^{(3)}$ are functionally independent and in involution in $C^\infty\bigr( (\mathbb{R}^2)^3 \bigl)$ with respect to (\ref{to}), that is,
\be
\bigl\{ F^{(2)}, h_{i}^{(3)}  \bigr\}_\omega= \bigl\{ F^{(3)}, h_{i}^{(3)}  \bigr\}_\omega=0,\qquad  
\bigl\{ F^{(2)},  F^{(3)} \bigr\}_\omega  =0 .
\label{tp2}
\ee
We observe in particular that 
the function $F=C\left(h^{(1)}_0,h^{(1)}_1,h^{(1)}_2,h^{(1)}_3\right)$ vanishes identically (see (\ref{ak})). Two additional constants of the motion can be deduced taking into account the permutation symmetry of the indices in $F^{(2)}$  (see  (\ref{am})):
  \be 
    F_{13}^{(2)}=S_{13} \bigl( F^{(2)}   \bigr)=(x_{3}-x_{2})   (y_{3}-y_{2}) ,\qquad 
      F_{23}^{(2)}=S_{23} \bigl( F^{(2)}   \bigr)=(x_{1}-x_{3})   (y_{1}-y_{3}) ,
\label{tq}
\ee
so that $F^{(3)}= F^{(2)}+ F_{13}^{(2)}+ F_{23}^{(2)}$.

The formalism thus provides us with four constants of the motion (\ref{tp}) and (\ref{tq}) for the oscillator Hamiltonian (\ref{tg}), which can be identified with some of the significant constants $k_i$ within the superposition rule  (\ref{LS01}) that we are looking for. We set 
\be
F^{(2)}=  k_1,\qquad  F_{23}^{(2)}=k_2,\qquad  F_{13}^{(2)}=k_3 ,\qquad F^{(3)}=F^{(2)}+F_{23}^{(2)}+F_{13}^{(2)}=  k_1+k_2+k_3\equiv k.
\label{tr}
\ee
There are several alternatives for the choice of the two necessary constants $k_i$ in order to obtain a superposition rule~\cite{Ballesteros6}. For computational reasons, we make the choice $F^{(2)}=k_1$ and $F^{(3)}=k$, in order to express   $(x_1,y_1)$ in terms of $(x_2,y_2,x_3,y_3)$,  along with the  two constants $k_1,k$. After some routine algebraic manipulations, we arrive at the following nonlinear superposition rule:
  \begin{equation}
\begin{split}
x_1 &  = x_3  +\frac{k-2k_1\pm B}{2(y_2-y_3)}  ,  \\[2pt]
y_1 &=y_3 +\frac{k-2k_1\mp B}{2(x_2-x_3)}  ,  
\end{split}
 \label{ts}
\end{equation}
where
\be
B= \sqrt{ \bigl( k- 2 (k_1+k_3)\bigr)^2  - 4 k_1 k_3 } \, ,\qquad k_3= (x_{3}-x_{2})   (y_{3}-y_{2}) .
 \label{tt}
\ee
 
 Finally, we apply the canonical transformation (\ref{sk}) to the above results while keeping the canonical symplectic form (\ref{to}),
    \be
\omega = \dd x_1\wedge \dd y_1+\dd x_2\wedge \dd y_2+\dd x_3\wedge \dd y_3 = \dd q_1\wedge \dd p_1+\dd q_2\wedge \dd p_2+\dd q_3\wedge \dd p_3.
\label{tu}
\ee
As a result, we get $t$-independent constants of the motion  (\ref{tp}) for the oscillator SIS Hamiltonian (\ref{tg}) expressed in terms of the  Hamiltonian vector fields 
 (\ref{th}) with  the canonical  SIS variables (\ref{z1}):
  \begin{equation}
\begin{split}
F^{(2)}&= \left( \frac {q_1^2 p_1^2-1} {p_1}- \frac {q_2^2 p_2^2-1} {p_2} \right)   \left( \frac{q_1p_1^2}{q_1^2p_1^2-1} -\frac{q_2p_2^2}{q_2^2p_2^2-1} \right) ,  \\[4pt]
F^{(3)}&=\sum_{1=i< j}^{3}  \left( \frac {q_i^2 p_i^2-1} {p_i}- \frac {q_j^2 p_j^2-1} {p_j} \right)   \left( \frac{q_ip_i^2}{q_i^2p_i^2-1} -\frac{q_jp_j^2}{q_j^2p_j^2-1} \right)  .  
\end{split}
 \label{tv}
\end{equation}
The corresponding superposition rule for (\ref{ti}) turns out to be
\begin{equation}
\begin{split}
q_1&=  { \left( \frac {q_3^2 p_3^2-1} {p_3}  +\frac{k-2k_1\pm B}{2\left(  \frac{q_2p_2^2}{q_2^2p_2^2-1}- \frac{q_3p_3^2}{q_3^2p_3^2-1}\right)}  \right)^2  \left( \frac{q_3p_3^2}{q_3^2p_3^2-1} +\frac{k-2k_1\mp B}{2 \left(\frac {q_2^2 p_2^2-1}{ p_2}-\frac {q_3^2 p_3^2-1} {p_3} \right)} \right)  } \\[2pt]
&\quad \times \left\{
{ \left(  \frac {q_3^2 p_3^2-1} {p_3}  +\frac{k-2k_1\pm B}{2\left(  \frac{q_2p_2^2}{q_2^2p_2^2-1}- \frac{q_3p_3^2}{q_3^2p_3^2-1} \right)}  \right)^2   \left( \frac{q_3p_3^2}{q_3^2p_3^2-1} +\frac{k-2k_1\mp B}{2\left(\frac {q_2^2 p_2^2-1}{ p_2}-\frac {q_3^2 p_3^2-1} {p_3} \right)} \right) ^2 -1} \right\}^{-1} \!, \\[4pt]
p_1&=\left\{ { \left(  \frac {q_3^2 p_3^2-1} {p_3}  +\frac{k-2k_1\pm B}{2\left(  \frac{q_2p_2^2}{q_2^2p_2^2-1}- \frac{q_3p_3^2}{q_3^2p_3^2-1} \right)}  \right)^2   \left( \frac{q_3p_3^2}{q_3^2p_3^2-1} +\frac{k-2k_1\mp B}{2\left(\frac {q_2^2 p_2^2-1}{ p_2}-\frac {q_3^2 p_3^2-1} {p_3} \right)} \right) ^2 -1}  \right\} \\
&\quad \times
{ \left(  \frac {q_3^2 p_3^2-1} {p_3}  +\frac{k-2k_1\pm B}{2\left(  \frac{q_2p_2^2}{q_2^2p_2^2-1}- \frac{q_3p_3^2}{q_3^2p_3^2-1}\right)}  \right)^{-1}} \, ,
\end{split}
 \label{tw}
\end{equation}
with $B$ given in (\ref{tt}) and 
\be
 k_3= \left( \frac {q_3^2 p_3^2-1} {p_3}- \frac {q_2^2 p_2^2-1} {p_2} \right)   \left( \frac{q_3p_3^2}{q_3^2p_3^2-1} -\frac{q_2p_2^2}{q_2^2p_2^2-1} \right).
 \label{tx}
\ee
Summarizing, we have obtained the general solution $(q_1,p_1)$ of the oscillator system of differential equations (\ref{ti}) generalizing the Hamiltonian (\ref{hamcovid}) in terms of two particular solutions  $(q_2,p_2)$ and  $(q_3,p_3)$ and two significant constants $k_1$ and $k$   in an explicit form. We remark that the superposition rule (\ref{tw}) holds for any $t$-dependent parameters  $b_1(t)$,  $b_2(t)$ and $ \rho_0(t)$ within the oscillator SIS Hamiltonian (\ref{tg}), so that this result can also be regarded also as a  superposition rule  for the book SIS system (\ref{sf}) as the  particular case with $b_1(t)=0$.


\sect{Generalized Hamiltonian from the two-photon algebra}
 \label{s5}

Following the classification of LH systems~\cite{LH2015,BHLS} on the plane $\mathbb R^2$, the embedding $\mathfrak{b}_2\subset \mathfrak{h}_4$ can still be extended up to a maximal chain $\mathfrak{b}_2\subset \mathfrak{h}_4\subset \mathfrak{h}_6$, with $\mathfrak{h}_6$ (class P$_5$ in~\cite{LH2015}) being the so-called two-photon algebra~\cite{Gilmore,BBF09}, that corresponds to the highest dimensional non-solvable LH algebra on the plane. This embedding of the oscillator algebra hence leads to a maximal extension of the book SIS system through this chain. Specifically, we have to add two vector fields $ \>X_4$ and  $\>X_5$ to the oscillator basis 
(\ref{ta}), which in the coordinates $(x,y)$ read
 \be
 \>X_1=\frac{\partial}{\partial x} ,\quad\  \>X_2\equiv    \>X_B=\frac{\partial}{\partial y},\quad\  \>X_3\equiv \>X_A=x\, \frac{\partial}{\partial x}- y\, \frac{\partial}{\partial y} , \quad\  \>X_4= y\, \frac{\partial}{\partial x} , \quad\  \>X_5= x\, \frac{\partial}{\partial y} .
      \label{ua}   
\ee
These generators satisfy the following commutation rules
\be
\begin{array}{llll}
[ \>X_1, \>X_2] = 0, &\quad [ \>X_1, \>X_3] = \>X_1,&\quad [ \>X_1, \>X_4] =0,&\quad [ \>X_1, \>X_5] = \>X_2,\\[2pt]
[ \>X_2, \>X_3] =- \>X_2,&\quad[ \>X_2, \>X_4] = \>X_1,&\quad [ \>X_2, \>X_5] =0,&\quad
[ \>X_3, \>X_4] =-2 \>X_4, \\[2pt]
[ \>X_3, \>X_5] =2 \>X_5,&\quad
[ \>X_4, \>X_5] =- \>X_3 ,&\quad   &
\end{array}
\label{ub}   
\ee
determining a 5-dimensional Lie algebra isomorphic to  $\mathfrak{sl}(2 ) \ltimes \mathbb{R}^2$ where $\mathfrak{sl}(2 )= \spn\{  \>X_3, \>X_4, \>X_5\}$ and 
$\mathbb{R}^2= \spn\{ \>X_1, \>X_2 \}$.  The corresponding   Lie system $\>X$   is now determined by the $t$-dependent vector field
\be
{\bf X} =  b_1(t) {\bf X}_1 +  b_2(t) {\bf X}_2+ \rho_0(t) {\bf X}_3+ b_4(t) {\bf X}_4 +  b_5(t) {\bf X}_5 ,
\label{uc} 
\ee
 giving rise to    the system of differential equations
\be
\begin{split}
\frac{\dd x}{\dd t}&= \rho_0(t) x+b_1(t) +b_4(t)y , \\[2pt]
 \frac{\dd y}{\dd t}&=- \rho_0(t)   y + b_2(t) +b_5(t)x.
 \end{split}
\label{ud}
\ee
Therefore, these equations generalize the oscillator Lie system (\ref{td}). Albeit the system is still linear for    $b_4(t)  b_5(t)\ne 0 $, an explicit integration is quite cumbersome and from the computational point of view, rather ineffective. Explicitly, from the first equation in (\ref{ud}) we find that
\be
y=\frac{1}{b_4(t)} \left(\frac{\dd x}{\dd t}- \rho_0(t) x- b_1(t)  \right) ,
\ee 
which leads  to the second-order inhomogeneous linear equation 
\begin{equation}\label{2or}
\frac{\dd^2 x}{\dd t^2}-\frac{\dd}{\dd t}(\log b_4(t))\frac{\dd x}{\dd t} + A(t) x= B(t),
\end{equation}
where 
\be
\begin{split}
A(t)& =  \rho_0(t)\frac{\dd}{\dd t}(\log b_4(t))-\rho_0^2(t)-b_4(t)b_5(t)-\frac{\dd \rho_0}{\dd t},\\
B(t) &= - b_1(t)\frac{\dd}{\dd t}(\log b_4(t))+\rho_0(t)b_1(t)+b_2(t)b_4(t)+\frac{\dd b_1}{\dd t}.
\end{split}
\ee
In order to solve the equation, a particular solution of the inhomogeneous equation and the general solution of the homogeneous part must be computed. The homogeneous part itself can be reduced by the change of variables 
\be
u=\frac 1{x} \,\frac{\dd x}{\dd t}
\ee
 to a Riccati equation 
\begin{equation}
\frac{\dd u}{\dd t}+u^2-\frac{\dd}{\dd t}(\log b_4(t))u +A(t)=0,
\end{equation}
which in general cannot be solved explicitly (i.e., the solution may not be expressible in terms of the classical functions). Alternatively, the equation (\ref{2or}) can be reduced to a free second-order equation $z^{\prime\prime}(\tau)=0$ using the Lie symmetry method \cite{BC}. However, the general solution is still far from being easily obtained, due to the complicated form of the linearizing point transformation. These computational difficulties justify the use of a superposition principle for the system (\ref{ud}), with the additional advantage that the results can be easily translated to the coordinates $(q,p)$ via the canonical transformation (\ref{sk}). We further observe that the system in these coordinates is far from being linear.

Starting from the realization (\ref{ua}), it can be easily verified that the generators $\>X_i$  are Hamiltonian vector fields $h_i$  with respect to the symplectic form (\ref{si}),  with functions  
\be
h_1=y,\qquad h_2\equiv h_B=-x,\qquad h_3\equiv h_A= x y,\qquad  h_4=\tfrac 12   y^2 ,\qquad  h_5=-\tfrac 12 x^2, \qquad  h_0 =1.
\label{ue}
\ee
Again, a central generator $h_0$ must be added to obtain a  Lie--Poisson algebra, where the Poisson brackets are given by
\be
\begin{array}{llll}
\{h_1,h_2\}_\omega=h_0, &\quad \{h_1,h_3\}_\omega=-h_1,&\quad \{h_1,h_4\}_\omega=0,&\quad \{h_1,h_5\}_\omega=-h_2,\\[2pt]
\{h_2,h_3\}_\omega=h_2,&\quad\{h_2,h_4\}_\omega=-h_1,&\quad \{h_2,h_5\}_\omega=0,&\quad
\{h_3,h_4\}_\omega=2h_4, \\[2pt]
\{h_3,h_5\}_\omega=-2h_5,&\quad
\{h_4,h_5\}_\omega=h_3, &\quad   \{h_0,\cdot\}_\omega=0 . &
\end{array}
\label{uf}
\ee
The resulting 6-dimensional LH algebra is a central extension of  $\mathfrak{sl}(2 ) \ltimes \mathbb{R}^2$, isomorphic 
to  the two-photon Lie algebra $\mathfrak{h}_6$~\cite{Gilmore,BBF09}, and further equivalent to the  (1+1)-dimensional centrally extended Schr\"odinger  Lie algebra~\cite{Schrod}. Clearly $\mathfrak{h}_6$ contains 
 the book    $\mathfrak{b}_2=\spn\{ h_2,h_3\}$ and   the oscillator   $\mathfrak{h}_4=\spn\{h_1, h_2,h_3,h_0\}$ LH algebras, as well as the simple Lie algebra $\mathfrak{sl}(2 )= \spn\{    h_3,h_4,h_5\} $. Hence we have the inclusions
\be
\mathfrak{b}_2\subset \mathfrak{h}_4\subset \mathfrak{h}_6,\qquad \mathfrak{sl}(2 )\subset \mathfrak{h}_6.
\label{ug}
\ee
 
The same  system (\ref{ud}) of differential equations can alternatively be deduced from the Hamilton equations with canonical variables $(x,y)$ using the Hamiltonian  associated to the vector field ${\bf X} $  (\ref{uc}):
   \begin{equation}
h= b_1(t)h_1 +  b_2(t) h_2+ \rho_0(t)h_3 +  b_4(t) h_4 +  b_5(t) h_5   .
 \label{uh}
\end{equation}

In analogy to the previous section, the canonical transformation (\ref{sk}) leads to the expression of the vector fields (\ref{ua})  and the  Hamiltonian vector fields (\ref{ue})  in  terms of the canonical variables $(q,p)$ (\ref{z1}), namely 
 \be
\begin{split}
 \>X_1&= - \frac{2qp}{(q^2p^2-1)^2} \,\frac{\partial}{\partial q}+\frac{p^2(q^2p^2+1)}{(q^2p^2-1)^2}\,\frac{\partial}{\partial p}, \qquad \>X_2\equiv   {\bf X}_B=-\left(q^2+\frac{1}{p^2}\right)\frac{\partial}{\partial q}+2 q p\,\frac{\partial}{\partial p},    \\[2pt]
    \>X_3&\equiv  {\bf X}_A=q\,\frac{\partial}{\partial q}-p\,\frac{\partial}{\partial p},\qquad   \>X_4= - \frac{2 q^2 p^3}{(q^2p^2-1)^3} \,\frac{\partial}{\partial q} +\frac{q p^4(q^2p^2+1)}{(q^2p^2-1)^3}  \,\frac{\partial}{\partial p} ,   \\[2pt]    \>X_5&= \frac{(1-q^4 p^4)}{p^3} \,\frac{\partial}{\partial q}  +  2 q (q^2p^2-1)   \,\frac{\partial}{\partial p},
   \\[2pt]
h_1&=\frac{qp^2}{q^2 p^2 -1},\qquad h_2\equiv h_B=\frac{1-q^2p^2}{p},\qquad h_3\equiv h_A= qp, \\[2pt]
  h_4&=\frac12\biggl(\frac{qp^2}{q^2 p^2 -1} \biggr)^2 ,\qquad    h_5=-\frac12\biggl(\frac{q^2p^2-1}{p}\biggr)^2 ,\qquad h_0 =1.
 \end{split}
\label{ui}
\ee
The corresponding first-order system that generalizes the oscillator SIS system (\ref{ti}) adds two new parameters $b_4(t)$ and $b_5(t)$:
\begin{equation} 
\begin{split}
\frac{\dd q}{\dd t}&=\rho_0(t) q-2b_1(t) \, \frac{qp}{(q^2p^2-1)^2}  - b_2(t) \left(  q^2+\frac{1}{p^2} \right) -2b_4(t) \, \frac{ q^2 p^3}{(q^2p^2-1)^3}  +b_5(t)\,\frac{(1-q^4 p^4)}{p^3} ,
\\[2pt]
\frac{\dd p}{\dd t}&=-\rho_0(t) p+b_1(t) \, \frac{p^2(q^2p^2+1)}{(q^2p^2-1)^2}  +2b_2(t)  qp + b_4(t)\, \frac{q p^4(q^2p^2+1)}{(q^2p^2-1)^3} + 2b_5(t)  q (q^2p^2-1)\, .
\end{split}
 \label{uj}
\end{equation}
This system is a formal generalization of the SIS epidemic time-dependent Hamiltonians associated to both the book and oscillator SIS systems (\ref{sf}) and (\ref{ti}). Whether this system truly corresponds to a realistic epidemic model (or other processes that can be modelled like contagions) depends on the fact whether the added parameters can be properly identified with some fluctuation parameters. Our purpose is not the analyze these possibilities, but rather to provide a procedure that allows to obtain solutions of such systems, using a superposition principle determined by the LH structure. 
It is worthy to be observed that, as a byproduct of the embeddings (\ref{ug}), this method also provides a $t$-dependent  system for the $\mathfrak{sl}(2)$-LH  algebra whenever  $b_1(t)= b_2(t)=0$, although this case cannot be considered as a realistic generalization of the SIS epidemic Hamiltonian of (\ref{sismodel3}) due to the vanishing of $b_2(t)$.

  
   \subsect{Superposition rule for the  two-photon time-dependent Hamiltonian}
 \label{s51}

In contrast to the previous  book and oscillator Hamiltonian systems, formerly written in Cartesian coordinates $(x,y)$ in (\ref{sl}) and (\ref{td}), the two-photon system (\ref{ud}) is nontrivially coupled, which prevents us from finding a solution by quadratures for the case $b_4(t)b_5(t)\neq 0$ (see equation (\ref{2or})). However, as already mentioned, we can deduce a superposition rule and then apply the canonical transformation   (\ref{sk}) to obtain a superposition rule for the system in the variables $(q,p)$ given by (\ref{z1}). To this extent, we follow the same steps as used in Section~\ref{s42} for the oscillator time-dependent Hamiltonian (the more technical details are described in the Appendix).

We start considering the  $\mathfrak{h}_6$-LH algebra in a basis $\{ v_1,\dots,v_5,v_0 \}$ that satisfies the  Poisson brackets
 (\ref{uf}). Besides the trivial central generator, the two-photon Lie--Poisson algebra possesses a third-order Casimir invariant given by (see~\cite{BHLS,BBF09}):
\be
C= 2\bigl(v_{1}^{2}v_{5}-v_{2}^{2}v_{4}-v_{1}v_{2}v_{3} \bigr) -v_{0}\bigl(v_{3}^{2}+4 v_{4}v_{5}\bigr) ,\qquad \{C ,\cdot\}=0 .
  \label{ul}
\ee
This implies that a superposition rule will be given in terms of $m=3$ particular solutions and $n=2$ significant constants, hence the relevant
indices for the application of the our formalism (see (\ref{aj})) are $k = 2, 3,4$. The Hamiltonian vector fields  $h^{(\kk)}_i$ are easily obtained from  (\ref{ue}), being explicitly given by 
  \begin{equation}
\begin{split}
  h^{(k)}_1 &=\sum_{\ss=1}^k y_s ,\qquad    h^{(k)}_2  =-\sum_{\ss=1}^k x_s,\qquad   h^{(k)}_3  =\sum_{\ss=1}^k x_s y_s ,    \\[2pt]
  h^{(k)}_4 &= \frac 12 \sum_{\ss=1}^k y^2_s   ,\qquad    h^{(k)}_5  =- \frac 12 \sum_{\ss=1}^k x^2_s,\qquad   h^{(k)}_0  =k,
\qquad k=2,3,4.\end{split}
 \label{um}
\end{equation}
These functions satisfy the 
 Poisson brackets    (\ref{uf}) with respect to the symplectic form
   \be
\omega = \dd x_1\wedge \dd y_1+\dd x_2\wedge \dd y_2+\dd x_3\wedge \dd y_3+\dd x_4\wedge \dd y_4 .
\label{un}
\ee
Now, introducing (\ref{um}) into the Casimir (\ref{ul}), we find $t$-independent constants of the motion  (\ref{ak}) for the diagonal prolongation $\widetilde{ \>X}^4$   to $(\mathbb{R}^2)^4$. These are $F=F^{(2)}=0$ and 
 \begin{equation}
\begin{split}
 F^{(3)}&=  
 \bigl(
x_{1}(y_{2}-y_{3})+x_{2}(y_{3}-y_{1})+x_{3}(y_{1}-y_{2}) \bigr)^{2}  , \\[2pt]
  F_{34}^{(3)}&=  \bigl(
x_{1}(y_{2}-y_{4})+x_{2}(y_{4}-y_{1})+x_{4}(y_{1}-y_{2}) \bigr)^{2}     , \\[2pt]
 F_{24}^{(3)}&=   \bigl(
x_{1}(y_{3}-y_{4})+x_{3}(y_{4}-y_{1})+x_{4}(y_{1}-y_{3}) \bigr)^{2}    , \\[2pt]
  F_{14}^{(3)}&=  \bigl(
x_{2}(y_{3}-y_{4})+x_{3}(y_{4}-y_{2})+x_{4}(y_{2}-y_{3}) \bigr)^{2}   , \\[2pt]
F^{(4)}&= F^{(3)}+   F_{34}^{(3)}+F_{24}^{(3)}+F_{14}^{(3)}  ,  
\end{split}
 \label{uo}
\end{equation}
where we have made use of the permutation symmetry  of the variables for $ F^{(3)}$.  

In this case, a convenient choice for the $n=2$ significant constants intervening in the superposition rule is given by the functions 
$ F^{(3)}=k_1^2$ and $  F_{34}^{(3)}=k_2^2$, where $k_1$ and $k_2$ are constants. In terms of these, after some algebraic manipulation, we can write $(x_1,y_1)$ as the general solution of the system (\ref{ud}) in terms of three particular solutions  $(x_2,x_2)$,  $(x_3,x_3)$ and $(x_4,x_4)$, together with the two constants $k_1$ and $k_2$, in the following form
\be
\begin{split}
x_1 &  =\left( 1+\frac{k_2-k_1}{k_4}\right) x_2-\frac{k_2}{k_4}  \, x_3  +\frac{k_1}{k_4}  \, x_4 , \\[2pt]
y_1 &=\left( 1+\frac{k_2-k_1}{k_4}\right) y_2-\frac{k_2}{k_4}  \, y_3  +\frac{k_1}{k_4}  \, y_4  ,  
  \end{split}
 \label{up}
\ee
where, in shorthand notation,
\be
 k_4 =  
x_{2}(y_{3}-y_{4})+x_{3}(y_{4}-y_{2})+x_{4}(y_{2}-y_{3}).
 \label{uq}
\ee
At this stage we apply the canonical transformation (\ref{sk}) to the above results, always preserving  the  symplectic form (\ref{un}),
    \be
\omega =\sum_{\ss=1}^4  \dd x_s\wedge \dd y_s  =\sum_{\ss=1}^4 \dd q_s\wedge \dd p_s .
\label{ur}
\ee
Then we directly deduce from (\ref{uo})  the $t$-independent constants of the motion in  the  canonical variables (\ref{z1}). In particular, the constant $k_1$ such that $F^{(3)}=k_1^2$   now becomes
 \be
  \begin{split}
k_1&=  
 \frac {q_1^2 p_1^2-1} {p_1} \left(  \frac{q_2p_2^2}{q_2^2p_2^2-1} - \frac{q_3p_3^2}{q_3^2p_3^2-1}\right)+ \frac {q_2^2 p_2^2-1} {p_2}\left(  \frac{q_3p_3^2}{q_3^2p_3^2-1}-  \frac{q_1p_1^2}{q_1^2p_1^2-1}  \right)  \\[2pt]
&\qquad + \frac {q_3^2 p_3^2-1} {p_3} \left(  \frac{q_1p_1^2}{q_1^2p_1^2-1}  -  \frac{q_2p_2^2}{q_2^2p_2^2-1}\right)   .  
\end{split}
 \label{us}
\end{equation}
The resulting superposition rule  for the two-photon system (\ref{uj}), although cumbersome, determines the general solution $(q_1,p_1)$ in terms of rational functions of the three particular solutions  $(q_2,p_2)$, $(q_3,p_3)$  and  $(q_4,p_4)$ and the significant constants $k_1$ and $k_2$, being explicitly given by
\begin{equation}
\begin{split}
q_1&= \left(  \biggl( 1+\frac{k_2-k_1}{k_4}\biggr) \frac {q_2^2 p_2^2-1} {p_2} -\frac{k_2}{k_4}  \, \frac {q_3^2 p_3^2-1} {p_3}  +\frac{k_1}{k_4}  \, \frac {q_4^2 p_4^2-1} {p_4}     \right)^2 \\[2pt]
&\qquad \times  \left(  \biggl( 1+\frac{k_2-k_1}{k_4}\biggr) \frac{q_2p_2^2}{q_2^2p_2^2-1}  -\frac{k_2}{k_4}  \, \frac{q_3p_3^2}{q_3^2p_3^2-1}  +\frac{k_1}{k_4}  \, \frac{q_4p_4^2}{q_4^2p_4^2-1}    \right) \\[2pt]
&\qquad \times  \left\{   \left(  \biggl( 1+\frac{k_2-k_1}{k_4}\biggr) \frac {q_2^2 p_2^2-1} {p_2} -\frac{k_2}{k_4}  \, \frac {q_3^2 p_3^2-1} {p_3}  +\frac{k_1}{k_4}  \, \frac {q_4^2 p_4^2-1} {p_4}     \right)^2 \right.   \\ 
&\qquad\qquad  \times \left. \left(  \biggl( 1+\frac{k_2-k_1}{k_4}\biggr) \frac{q_2p_2^2}{q_2^2p_2^2-1}  -\frac{k_2}{k_4}  \, \frac{q_3p_3^2}{q_3^2p_3^2-1}  +\frac{k_1}{k_4}  \, \frac{q_4p_4^2}{q_4^2p_4^2-1}    \right)^2-1 \right\}^{-1} ,\\[4pt] 
p_1&=       \left\{   \left(  \biggl( 1+\frac{k_2-k_1}{k_4}\biggr) \frac {q_2^2 p_2^2-1} {p_2} -\frac{k_2}{k_4}  \, \frac {q_3^2 p_3^2-1} {p_3}  +\frac{k_1}{k_4}  \, \frac {q_4^2 p_4^2-1} {p_4}     \right)^2 \right.  \\ 
&\qquad\qquad  \times \left. \left(  \biggl( 1+\frac{k_2-k_1}{k_4}\biggr) \frac{q_2p_2^2}{q_2^2p_2^2-1}  -\frac{k_2}{k_4}  \, \frac{q_3p_3^2}{q_3^2p_3^2-1}  +\frac{k_1}{k_4}  \, \frac{q_4p_4^2}{q_4^2p_4^2-1}    \right)^2-1 \right\}  \\
&\qquad   \times  \left(  \biggl( 1+\frac{k_2-k_1}{k_4}\biggr) \frac {q_2^2 p_2^2-1} {p_2} -\frac{k_2}{k_4}  \, \frac {q_3^2 p_3^2-1} {p_3}  +\frac{k_1}{k_4}  \, \frac {q_4^2 p_4^2-1} {p_4}     \right)^{-1} ,
\end{split}
 \label{ut }
\end{equation}
where
 \be
  \begin{split}
k_4&=  
 \frac {q_2^2 p_2^2-1} {p_2} \left(  \frac{q_3p_3^2}{q_3^2p_3^2-1} - \frac{q_4p_4^2}{q_4^2p_4^2-1}\right)+ \frac {q_3^2 p_3^2-1} {p_3}\left(  \frac{q_4p_4^2}{q_4^2p_4^2-1}-  \frac{q_2p_2^2}{q_2^2p_2^2-1}  \right)  \\[2pt]
&\qquad + \frac {q_4^2 p_4^2-1} {p_4} \left(  \frac{q_2p_2^2}{q_2^2p_2^2-1}  -  \frac{q_3p_3^2}{q_3^2p_3^2-1}\right)  . 
\end{split}
 \label{uv}
\end{equation}
The relevant observation at this point is that this superposition principle remains unaltered for any choice of the $t$-dependent parameters $b_i(t)$  and $ \rho_0(t)$ appearing in the Hamiltonian (\ref{uh}), thus holds for all the Hamiltonian systems considered previously, and, in particular, for the SIS models (\ref{sismodel3}) and (\ref{sismodel3b}). This means specifically that if any of the generalized systems turns out to be experimentally identified with a realistic model, the main properties of the original systems remain unaltered.


\section{Concluding remarks}
\label{s6}

In the context of SIS epidemic models, the Hamiltonian formulation has shown to be an effective technique to unveil the 
main features concerning the dynamics of such systems, even in those cases where a closed expression of the solutions is not possible. One of such models~\cite{NakamuraMartinez} has recently been reconsidered from this point of view, which has further led to a generalization involving a time-dependent infection rate~\cite{Covid}. Both systems correspond to a same class of LH systems, allowing a unified approach. Taking into account that the LH algebra that underlies these models can be embedded into higher dimensional LH algebras on the real plane, in this work we have systematically analyzed the formal extensions of the SIS models to first-order systems of differential equations depending on additional time-dependent parameters. Moreover, it has been shown that any Hamiltonian system (whether identifiable with a SIS model or not) based on the book algebra  $\mathfrak{b}_2$ admits an exact solution in closed form, which particularized to the above mentioned models reproduce the exact solution given in~\cite{NakamuraMartinez} and completes the analysis in~\cite{Covid}, providing the general solution of the latter. The general $\mathfrak{b}_2$-LH system already generalizes formally these systems, as it introduces a second parameter, the interpretation of which has however not been considered. In addition, using the embedding chain $\mathfrak{b}_2\subset \mathfrak{h}_4\subset \mathfrak{h}_6$ of LH algebras~\cite{LH2015}, further formal extensions of the models have been proposed, up to a maximal case depending on five arbitrary time-dependent parameters. While the systems based on the oscillator  LH algebra $\mathfrak{h}_4$ have been shown to admit an exact solution (this case additionally completing the results in~\cite{Covid}, where a superposition rule was given but with non-canonical variables), the computational obstructions to effectively integrate the corresponding LH system for $\mathfrak{h}_6$ suggests to derive a superposition principle, the existence of which is guaranteed by the properties of LH systems. This case also represents the maximal possible extension of the time-dependent Hamiltonian models based on the assumptions of~\cite{NakamuraMartinez} and~\cite{Covid}. This means that, once the additional parameters have been identified with suitable characteristics of an epidemic model, the formalism developed in this work immediately provides either the general solution of the system, or a superposition rule that depends at most on three particular solutions.   

As mentioned, we have focused primarily on the analytical and geometrical properties of LH systems that contain the models (\ref{sismodel3b}), without pursuing further the viability of such models at the epidemiological or biological level. As the method developed allows to derive solutions whatever the particular significance of the parameters, for any appropriate epidemiological extension the main properties are directly obtained by insertion of the values of the parameters in the corresponding solutions. In this context, the relevant question that emerges towards realistic applications is whether these additional arbitrary time-dependent functions can effectively  identified with some compartmental parameters subjected to fluctuations, in the line of the first generalization proposed in~\cite{Covid}. In this sense, each possible choice of the parameters must be studied separately, with a proper contextualization of the parameters, based on experimental data. The epidemiological applicability of the generalized LH systems remains currently open.

It may be asked whether the LH formalism can be applied to other epidemic models, specifically in order to obtain exact solutions. The models considered in~\cite{Pra,bon} lead, for certain identifications of the parameters, to differential equations very similar to either the matrix or projective Riccati equations. Although these are known to correspond to Lie systems (with semisimple Vessiot--Guldberg algebra) and hence to admit a superposition rule (see e.g.~\cite{LuSa,Wi} and references therein), in general they do not admit an LH structure, and thus their integrability by quadratures must be analyzed by other means. Actually, among the 28 isomorphism classes of Lie algebras of vector fields over $\mathbb{R}^2$, only 12 of them correspond to algebras of Hamiltonian vector fields \cite{LH2015}, and further applications to models in biology or information theory should be searched among these LH algebras. Concerning the integrability by quadratures of LH systems, it can be inferred from the general theory of Lie systems, as long as the Vessiot--Guldberg algebra is solvable \cite{Car1,Car4}, while for the non-solvable case it is not true in general, although for certain Lie systems integrability conditions may be found that, combined with a reduction procedure \cite{Car5}, may lead to an integration by quadratures. It is currently an open problem whether, for the case of non-solvable Vessiot--Guldberg algebras, the compatibility with a symplectic structure, beyond providing a systematic procedure to construct superposition rules, also determines additional integrability conditions that allow to explicitly integrate a system.

 Although the extension of SIS models containing (\ref{sismodel3}) as a particular case has been exhausted via the LH formalism, there is an additional possibility of further extending the systems, by applying the so-called Poisson--Hopf  algebra deformation of LH systems formalism recently introduced in~\cite{Ballesteros6, BCFHLb, BCFHLc}, which is based on quantum groups~\cite{BBHMR09,CP}. The main idea behind this procedure is to combine LH systems with quantum deformations, leading to genuinely nonlinear systems that are albeit no more described by a finite-dimensional LH algebra, but in terms of involutive functional distributions. As shown in~\cite{
 Ballesteros6, BCFHLb, BCFHLc}, for such systems a kind of deformed superposition rule and deformed systems of differential equations can also be found, with the additional feature that the quantum deformation parameter  can be identified with a perturbation parameter. In this frame, it is conceivable to analyze the generic deformed systems resulting from the embedding chains (\ref{ug}) $\mathfrak{b}_2\subset \mathfrak{h}_4\subset \mathfrak{h}_6$ and $ \mathfrak{sl}(2 ) \subset \mathfrak{h}_6$ (a special case with a quantum deformation of $ \mathfrak{sl}(2)$  has already been considered in~\cite{Covid}), and study to which extent the corresponding equations can be explicitly solved. In this respect, we recall that all possible  quantum deformations for the oscillator Lie algebra  $ \mathfrak{h}_4$ can be found in~\cite{h4}, while those corresponding to the two-photon  $\mathfrak{h}_6$  were presented in~\cite{Schrod} in a   Schr\"odinger Lie algebra basis. Work along the various research lines commented above is currently in progress.
 

\section*{Acknowledgments}

\phantomsection
\addcontentsline{toc}{section}{Acknowledgments}

R.C.S.~and F.J.H.~have been partially supported by Agencia Estatal de Investigaci\'on (Spain)  under grant  PID2019-106802GB-I00/AEI/10.13039/501100011033.  F.J.H.~also acknowledges support  by the Regional Government of Castilla y Le\'on (Junta de Castilla y Le\'on, Spain) and by the Spanish Ministry of Science and Innovation MICIN and the European Union NextGenerationEU  (PRTR C17.I1). The authors acknowledge the referees for valuable comments and suggestions that have greatly improved the presentation, as well as for pointing out additional pertinent references.


\section*{Appendix. Basic properties of Lie--Hamilton systems}
\setcounter{equation}{0}
\renewcommand{\theequation}{A.\arabic{equation}}

  \phantomsection
\addcontentsline{toc}{section}{Appendix. Basic properties of Lie--Hamilton systems}

We briefly recall the main definitions and properties of  LH systems. For additional details   the reader is referred to~\cite{CLS13,BCHLS13Ham,LuSa,BHLS,LH2015} and references therein.
  
Given a system of first-order ordinary
differential equations on a manifold $M$ of dimension $n$ with coordinates $\>x= \{x_1,\dots, x_n \}$,
\begin{equation} 
\frac{{\rm d} x_j}{{\rm d}t}=f_j(t,\>x ), \qquad j=1,\dots, n,
 \label{LS0}
\end{equation}
we say that it admits a fundamental system of solutions if the general solution
can be written in terms of $m\leq n$ 
functionally independent particular solutions $ \{  {\bf
x}_{1},\dots,{\bf x}_{m} \}  $, where ${\bf x}_{s}= \bigl((x_{1})_{s},\dots,(x_{n})_{s}\bigr)$,  and $n$   constants
$\left\{ k_{1},\dots,k_{n}\right\}  $:
\begin{equation}
x_{j}=\Psi_{j} ( {\bf x}_{1},\dots,{\bf
x}_{m};k_{1},\dots,k_{n} ) , \qquad j=1,\dots, n .
\label{LS01}
\end{equation}
The
relation (\ref{LS01}) is called a superposition rule for the
system (\ref{LS0}), while  the set $ \{  {\bf
x}_{1},\dots,{\bf x}_{m} \}  $ provides   a fundamental system of solutions. It is straightforward to verify that this  system 
can be described in equivalent form by means of the vector
field 
\begin{equation}
{\mathbf X} (  t,\mathbf{x} )
=\sum_{j=1}^{n}f_{j} ( t,\mathbf{x} )
\frac{\partial}{\partial x_{j}}\, ,    
\label{TDEP}
\end{equation}
called the $t$-dependent vector field associated to the system (\ref{LS0}). The existence of fundamental
systems of solutions was first analyzed by Lie \cite{LSc}, who established  a characterization in terms 
of finite-dimensional Lie algebras. The Lie--Scheffers Theorem asserts that the system (\ref{LS0}) admits a fundamental system of
solutions if and only if 
it can be represented in terms of $\ell$ $t$-dependent parameters $b_i(t)$ in the form
\begin{equation}
\frac{{\rm d}x_{j}}{{\rm d}t}=\sum_{i=1}^{\ell}b_{i}(t)  
\xi_{ij} (\mathbf{x} ) ,  \qquad j=1,\dots, n  ,
 \label{LS1}
\end{equation}
and such that the vector fields
\begin{equation}
{\mathbf X}_{i}(\>x) =\sum_{j=1}^{n}\xi_{ij} (  \mathbf{x} )
\frac{\partial}{\partial
x_{j}} \, , \qquad i=1,\dots, \ell ,
\label{LS2}
\end{equation}
span an $\ell$-dimensional Lie algebra $\mathfrak{g}$, where the numerical constraint 
$n\,m\geq \ell=\dim\mathfrak{g}$  is satisfied. Thus the generators  (\ref{LS2}) obey  the generic commutation relations given by
\begin{equation}
[ {\mathbf X}_{a} ,{\mathbf X}_{b} ]= \sum_{{c=1}}^{\ell} C_{ab}^{c}{\mathbf X}_{c},\qquad  a,b=1,\dots, \ell,\qquad C_{ab}^{c}\in\mathbb{R}. 
\label{xa}
\end{equation}
In these conditions, the $t$-dependent
vector field (\ref{TDEP}) is reformulated as
\begin{equation}
{\mathbf X} (  t,{\bf x} )
=\sum_{i=1}^{\ell}b_{i}(t) {\mathbf X}_{i}(\>x) .
\label{TDEPA}
\end{equation}
The Lie algebra $\mathfrak{g}$ generated by the vector fields ${\mathbf X}_i (  {\bf x} )$ is usually called a
Vessiot--Guldberg Lie algebra of the system (\ref{LS0}), while either the system  itself or  ${\mathbf X} (  t,{\bf x} )$   are referred to as a Lie system~\cite{LuSa,PW}.\footnote{It should  be observed that Vessiot--Guldberg  algebras are not uniquely
determined, and hence do not constitute an invariant of the system. It has thus sense to speak about minimality, in the sense that $\mathfrak{g}$ is minimal if no proper subalgebra of $\mathfrak{g}$ is a Vessiot--Guldberg--Lie algebra of
the system (\ref{LS0}).}

   A Lie system ${\mathbf X} (  t,{\bf x} )$ is called an LH  system if it admits a Vessiot--Guldberg Lie algebra $\mathfrak{g}$ of Hamiltonian vector fields   with respect to a Poisson structure  \cite{LuSa}, with the compatibility condition of the generators ${\mathbf X}_i (  {\bf x} )$    and the symplectic form $\omega$ being given by the Lie derivative:
\begin{equation}
{\mathcal L}_{\mathbf{X}_i}\omega =0, \qquad i=1,\dots, \ell.
\label{ad}
\end{equation}
The corresponding Hamiltonian functions $h_i ({\bf x})$ associated to ${\bf X}_i({\bf x})$ are obtained through the inner product (see e.g.~\cite{ARN})
\begin{equation}
\iota_{{\bf X}_i}\omega={\rm d}h_i,  \qquad i=1,\dots, \ell.
\label{ae}
\end{equation}
The Hamiltonian functions $h_i({\bf x})$ span (eventually adjoining a constant function $h_0$) an $\ell$-dimensional Lie algebra ${\mathcal H}_\omega$ called an LH  algebra, with Poisson brackets given by 
\begin{equation}
\{h_a,h_b\}_{\omega}= -\sum_{{c=1}}^{\ell} C_{ab}^{c}h_{c},\qquad a,b=1,\dots, \ell,
\label{af}
\end{equation}
where $C_{ab}^{c}$ are the same structure constants written in (\ref{xa}). Therefore, $\mathcal{H}_{\omega}$ is spanned by a set of functions $\left\{h_1,\ldots, h_\ell\right\}\subset C^\infty(\mmm)$, where $\mmm$ is a suitable  submanifold of $M$ that ensures that each $h_i$ is well defined.
The    $t$-dependent  Hamiltonian function $h (  t,{\bf x} )$ associated with the $t$-dependent vector field (\ref{TDEPA}) leading to the same Lie system (\ref{LS0}), is given by 
\be
h (  t,{\bf x} )=\sum_{i=1}^\ell b_i(t)h_{i} (  {\bf x} ).
\label{ag}
\ee

As LH systems correspond to a particular class of Lie systems, a superposition rule (\ref{LS01}) is guaranteed. In general, standard methods~\cite{PW,CGM07} for   deriving superposition rules  for Lie systems  require the   integration of PDEs or ODEs, which is often a cumbersome task. Nevertheless, if the Lie system is also an LH one, there is an algebraic approach~\cite{BCHLS13Ham} (see also~\cite{BHLS,Ballesteros6}) that allows a systematic obtainment of $t$-independent constants of the motion, from which a superposition rule can be deduced. Such a formalism is based on the so-called coalgebra symmetry approach for superintegrable systems (see~\cite{BBHMR09} and references theirein for technical details). The only formal requirement is that the LH algebra possesses a non-trivial Casimir invariant.

Explicitly,  let us consider the  LH algebra ${\cal H}_\omega$ of a LH system $\>X$ with Hamiltonian (\ref{ag}), expressed as a Lie--Poisson algebra with generators $\{ v_1,\dots,v_\ell \}$ that formally satisfy the same Poisson brackets (\ref{af}), and assume that it admits a non-constant Casimir  function 
\be
C=C(v_1,\dots,v_\ell),\qquad \{C,v_i\}=0,\qquad i=1,\dots, \ell.
\label{ai}
\ee
The underlying coalgebra structure of the LH algebra leads to the definition of  the following Hamiltonian vector fields
\begin{equation}
\begin{split}
 h^{(1)}_i   & := h_i(\>x_1)\subset C^\infty(\mmm), \qquad i=1,\dots, \ell,\\[2pt]
 h^{(\kk)}_i  & :=  h_i(\>x_1)+\cdots + h_i(\>x_\kk)\subset C^\infty(\mmm^k) ,\qquad k=2,\dots, m+1 ,\end{split}
 \label{aj}
\end{equation}
where ${\bf x}_{s}= \bigl((x_{1})_{s},\dots,(x_{n})_{s}\bigr)$ denotes the coordinates in the  $s^{th}$-copy submanifold $\mmm\subset M$ within  the product manifold $\mathcal{M}^k$. It can be proved (see~\cite{BCHLS13Ham,BHLS,Ballesteros6}) that the functions defined through the Casimir (\ref{ai})  and the Hamiltonian vector fields (\ref{aj}) by
\begin{equation}
  F:=C\left( h^{(1)}_1,\dots, h^{(1)}_\ell\right) ,\qquad F^{(\kk)} :=  C\left(h^{(\kk)}_1,\dots,h^{(\kk)}_\ell\right) ,   \qquad \kk=2,\ldots,m+1,
  \label{ak}
\end{equation}
are  $t$-independent constants of the motion for the diagonal prolongation
$\widetilde {\>X}^{m+1}$  of the LH system $\>X$  to the product manifold $\mmm^{m+1}$, with $\widetilde {\>X}^{m+1}$ given by
\begin{equation}
\widetilde{\>X}^{m+1}(t,\>x_1,\ldots,\>x_{m+1}):=
\sum_{k=1}^{m+1}\sum_{j=1}^nf_j(t,\>x_k)\frac{\partial }{\partial x^j}=\sum_{i=1}^\ell b_i(t)\>X_{ h^{(m+1)}_i} \, .
\label{al}
\end{equation}
Observe that the $ F^{(k)} $ are functions of $C^\infty(\mmm^{m+1})$, and that they can be considered as $t$-independent  constants of the motion for the  LH system   (\ref{ag}).  Furthermore,  if all the $F^{(k)}$ are non-constant,    they constitute a set of  $m$  functionally independent functions in $C^\infty(\mmm^{m+1})$ that are in involution (i.e., they Poisson commute). Moreover, the functions $F^{(k)}$ can be used to generate other 
 constants of the motion   in the form
\begin{equation}
F_{ij}^{(k)}=S_{ij} \bigl( F^{(k)}   \bigr) , \qquad 1\le  i<j\le  m+1,
\label{am}
\end{equation}
where $S_{ij}$ denotes the permutation of the variables $\>x_i\leftrightarrow
\>x_j$ in $\mmm^{m+1}$. 
  
We finally recall that, in order to obtain a superposition rule (\ref{LS01}) that depends on $m$ particular solutions, one should find a set  $\{I_1,\ldots,I_n\}$ of $t$-independent constants of the motion on $\mathcal{M}^{m+1}$ for $\widetilde{\> X}^{m+1}$ such that~\cite{LuSa}
\be
\frac{\partial (I_1,\ldots,I_n) }{\partial ( (x_1)_{m+1},\dots, (x_n)_{m+1}   )}\neq 0.
\label{an}
\ee
 This allows us to express the coordinates  $\>x_{m+1}=\left\{ (x_1)_{m+1},\dots, (x_n)_{m+1}\right\}$     in terms of the remaining coordinates in $\mathcal{M}^{m+1}$ and the constants $k_1,\ldots,k_n$   defined by  the conditions $I_1=k_1,\ldots,I_n=k_n$.
The constants of the motion  (\ref{ak}) and  (\ref{am}) are generally sufficient to determine the set $\{I_1,\ldots,I_n\}$, and hence to 
 deduce a superposition rule for the LH system in a direct algebraic way.
 


\footnotesize

 \begin{thebibliography}{99}

\phantomsection
\addcontentsline{toc}{section}{References}


\bibitem{Bai} N.~T.~J.~Bailey. {\it The Mathematical Theory of Infectious Diseases and its Applications}. (London: Griffin) 1975.

\bibitem{Hethcote}
H.~W.~Hethcote. {Three basic epidemiological models}. In: Levin S.A., Hallam T.G. and Gross L.J. (eds.),
  {\em Applied Mathematical Ecology. Biomathematics} vol.~18, (Berlin: Springer) 1989, pp.~119--144.  
 \newblock \href {https://doi.org/10.1007/978-3-642-61317-3_5}
  {\path{doi:10.1007/978-3-642-61317-3_5}}
 
\bibitem{Walt} F.~Hoppensteadt and P.~Waltman. A problem in the theory of epidemics. {\it Math. Biosci.} {\bf 9} (1970) 71--91.   \newblock \href {https://doi.org/10.1016/0025-5564(70)90094-5}
  {\path{doi:10.1016/0025-5564(70)90094-5}}

 \bibitem{KK}
W.~O.~Kermack and A.~G.~McKendrick. {A contribution to the mathematical theory of epidemics}. {\em Proc. R. Soc. Lond. A} {\bf 772} (1927) 700--721. 
\href {https://doi.org/10.1098/rspa.1927.0118}
  {\path{doi:10.1098/rspa.1927.0118}}

\bibitem{Ab52}
H.~Abbey. {An examination of the Reed-Frost theory of epidemics}. {\it Hum. Biol.} {\bf 24} (1952) 201--233. 

 \bibitem{Miller}
J.~C.~Miller. {Mathematical models of SIR disease spread with combined non-sexual and sexual transmission routes}. {\it Infect. Dis. Model.} {\bf 2} (2017) 35--55. 
\href {https://doi.org/10.1016/j.idm.2016.12.003}
  {\path{doi:10.1016/j.idm.2016.12.003}}

\bibitem{NakamuraMartinez}
G.~M.~Nakamura and A.~S.~Martinez.
Hamiltonian dynamics of the SIS epidemic model with stochastic fluctuations.
{\em Sci. Rep.} {\bf 9} (2019) 15841. 
\href {https://doi.org/10.1038/s41598-019-52351-x}
  {\path{doi:10.1038/s41598-019-52351-x}}

\bibitem{Bar} M.~S.~Bartlett. {\it Stochastic Population Models in Ecology and Epidemiology}. (London: Methuen) 1960.

\bibitem{Bun} H.~Bunke. {\it Gew\"ohnliche Differentialgleichungen mit zuf\"alligen Parametern}.  (Berlin:  Akademie-Verlag) 1972.

\bibitem{Or3} J.~A.~L\'azaro-Cam\'{\i} and J.~P.~Ortega. The stochastic Hamilton-Jacobi equation. {\it J. Geom. Mech.} {\bf 1} (2009) 295--315.
\href {https://doi.org/10.3934/jgm.2009.1.295}
 {\path{doi.org/10.3934/jgm.2009.1.295}}   
 
\bibitem{NuLe04}
 M.~C.~Nucci and P.~G.~L.~Leach. An integrable SIS model.  {\it J. Math. Anal. Appl.} {\bf 290} (2004) 506--518. 
 \href {https://doi.org/10.1016/j.jmaa.2003.10.044}
  {\path{doi:10.1016/j.jmaa.2003.10.044}}

\bibitem{Ba2020}
A.~Ballesteros, A.~Blasco and I.~Gutierrez-Sagredo. Hamiltonian structure of compartmental epidemiological models. {\em Physica D} {\bf 413}  (2020) 132656. 
\href {https://doi.org/10.1016/j.physd.2020.132656}
  {\path{doi:10.1016/j.physd.2020.132656}}
  
\bibitem{Covid} O.~Esen, E.~Fern\'andez-Saiz, C.~Sard\'on and M.~Zajac.  A generalization of a SIS epidemic model with fluctuations. {\em Math.~Meth.~Appl.~Sci.} {\bf 45}  (2022) 3718--3731. 
 \href {https://doi.org/10.1002/mma.8013}
  {\path{doi:10.1002/mma.8013}}

\bibitem{Bohner2019}
M.~Bohner, S.~Streipert and D.~F.~M.~Torres.  {Exact solution to a dynamic SIR model}. {\em Nonlinear Anal.: Hybrid Syst.} {\bf 32}  (2019) 228--238. 
\href {https://doi.org/10.1016/j.nahs.2018.12.005}
 {\path{doi:10.1016/j.nahs.2018.12.005}}

\bibitem{Ba2021}
A.~Ballesteros, A.~Blasco and I.~Gutierrez-Sagredo. Exact closed-form solution of a modified SIR model.  {\em Preprint  	arXiv:2007.16069} (2020). 
\href {https://doi.org/10.48550/arXiv.2007.16069}
  {\path{doi:10.48550/arXiv.2007.16069}}

  \bibitem{Kopfova1}
Z.~Chladn\'a, J.~Kopfov\'a, D.~Rachinskii and S.~C.~Rouf. Global dynamics of SIR model with switched transmission rate. 
{\em J. Math. Biol.}  {\bf 80} (2020) 1209--1233. 
\href {https://doi.org/10.1007/s00285-019-01460-2}
  {\path{doi:10.1007/s00285-019-01460-2}}

\bibitem{Kopfova2}
Z.~Chladn\'a, J.~Kopfov\'a, D.~Rachinskii and P.~\v Step\'anek. Effect of quarantine strategies in a compartmental model with asymptomatic groups. 
{\em J. Dyn. Diff. Equat.}    (2021). 
\href {https://doi.org/10.1007/s10884-021-10059-5}
  {\path{doi:10.1007/s10884-021-10059-5}}
  
\bibitem{Or1} J.~A.~L\'azaro-Cam\'{\i} and J.~P.~Ortega. Stochastic Hamiltonian dynamical systems. {\it
Rep. Math. Phys.} {\bf  61} (2008) 65--122.
\href {https://doi.org/10.1016/S0034-4877(08)80003-1}
 {\path{doi.org/10.1016/S0034-4877(08)80003-1}}  

\bibitem{Or2} J.~A.~L\'azaro-Cam\'{\i} and J.~P.~Ortega. Reduction, reconstruction, and skew-product
decomposition of symmetric stochastic differential equations. {\it Stoch. Dyn.} {\bf 9} (2009) 1--46.
\href {https://doi.org/}10.1142/S0219493709002531
 {\path{doi.org/10.1142/S0219493709002531}}  
 
 \bibitem{Pan} A.~Gray, D.~Greenhalgh, L.~Hu, X.~Mao and J.~Pan. A stochastic differential equation
SIS epidemic model. {\it SIAM J. Appl. Math.} {\bf 71} (2011) 876--902.
\href {https://doi.org/10.1137/10081856X}
 {\path{doi.org/10.1137/10081856X}}  

\bibitem{Otu} O.~M.~Otunuga. Time-dependent probability distribution for number of infection
in a stochastic SIS model: case study COVID-19. {\it Chaos, Solitons \& Fractals} {\bf 147} (2021)
110983.
\href {https://doi.org/10.1016/j.chaos.2021.110983}
{\path{doi.org/10.1016/j.chaos.2021.110983}}     

\bibitem{groh} J.~Groh. A stochastic differential equation for a class of Feller's one-dimensional
diffusion. {\it Math. Nachr.} {\bf 107} (1982) 267--271.   
\href {https://doi.org/10.1002/mana.19821070122}
 {\path{doi.org/10.1002/mana.19821070122}}     
 
\bibitem{CLS13}
J.~F.~Cari{\~n}ena, J.~de Lucas and C.~Sard{\'o}n.
{Lie--Hamilton systems: theory and applications}.
{\em Int. J. Geom. Methods Mod. Phys.} {\bf 10} (2013) 1350047.
  \href {https://doi.org/10.1142/S0219887813500473}
  {\path{doi:110.1142/S0219887813500473}}

\bibitem{BCHLS13Ham}
A.~Ballesteros, J.~F.~Cari{\~n}ena, F.~J.~Herranz, J.~de Lucas and C.~Sard{\'o}n. From constants of motion to superposition rules for Lie--{H}amilton systems.  {\it J. Phys. A: Math. Theor.} {\bf 46} (2013) 285203. 
\href {https://doi.org/10.1088/1751-8113/46/28/285203}
  {\path{doi:10.1088/1751-8113/46/28/285203}}

\bibitem{LH2015}
A.~Ballesteros, A.~Blasco, F.~J.~Herranz, J.~de Lucas and  C.~Sard{\'o}n. Lie--Hamilton systems on the plane: Properties, classification and applications.
 {\it J. Diff. Equ.} {\bf 258} (2015) 2873--2907.
 \href {https://doi.org/10.1016/j.jde.2014.12.031}
  {\path{doi:10.1016/j.jde.2014.12.031}}

\bibitem{BHLS} A.~Blasco, F.~J.~Herranz, J.~de Lucas and C.~Sard\'on.
{Lie--Hamilton systems on the plane: Applications and superposition rules.}
 {\it J. Phys. A: Math. Theor.}  {\bf 48} (2015)  345202. 
  \href {https://doi.org/10.1088/1751-8113/48/34/345202}
  {\path{doi:10.1088/1751-8113/48/34/345202}}

\bibitem{LuSa}
J.~de Lucas and  C.~Sard{\'o}n.
\emph{A Guide to Lie Systems with Compatible Geometric Structures}. (Singapore:
World Scientific)  2020.
\href {https://doi.org/10.1142/q0208}
  {\path{doi:10.1142/q02080}}

\bibitem{PW}
P.~Winternitz. 
{Lie groups and solutions of nonlinear differential equations}. In 
{\em Nonlinear Phenomena (Lectures Notes in Physics} {\bf{189}}),  K.B.~Wolf (ed.),
(New York: Springer) 1983, pp.~263--331.
 
\bibitem{CGM07} 
J.~F.~Cari{\~n}ena, J.~Grabowski and  G.~Marmo.  Superposition rules, Lie theorem and partial
differential equations.
 {\em Rep. Math. Phys.} {\bf 60} (2007)  237--258.
\href {https://doi.org/10.1016/S0034-4877(07)80137-6}
  {\path{doi:10.1016/S0034-4877(07)80137-6}}
  
\bibitem{Or4} J.~A.~L\'azaro-Cam\'{\i} and J.~P.~Ortega. Superposition rules and stochastic Lie--Scheffers systems. {\it Ann. Inst. Henri Poincar\'e Probab. Stat.} {\bf 45} (2009) 910--931.
\href {https://doi.org/10.1214/08-AIHP189}
 {\path{doi.org/10.1214/08-AIHP189}}       
  
\bibitem{ARN} V.~I.~Arnol'd. {\it Mathematical Methods of Classical Mechanics}. (New York: Springer) 1989.
\href {https://doi.org/10.1007/978-1-4757-2063-1}
  {\path{doi:10.1007/978-1-4757-2063-1}}

\bibitem{BBHMR09}
A.~Ballesteros, A.~Blasco, F.~J.~Herranz, F.~Musso and O.~Ragnisco, (Super)integrability from coalgebra symmetry: formalism and applications.
 {\it J. Phys. Conf. Ser.} {\bf 175} (2009) 012004. 
   \href {https://doi.org/10.1088/1742-6596/175/1/012004}
  {\path{doi:10.1088/1742-6596/175/1/012004}}
  
\bibitem{Ballesteros6}
A.~Ballesteros, R.~Campoamor-Stursberg, E.~Fern\'andez-Saiz, F.~J.~Herranz and J.~de Lucas.
{Poisson--Hopf deformations of Lie--Hamilton systems revisited: deformed superposition rules and applications to the oscillator algebra.}
{\it J. Phys. A: Math. Theor.} {\bf 54}  (2021) 205202. 
  \href {https://doi.org/10.1088/1751-8121/abf1db}
  {\path{doi:10.1088/1751-8121/abf1db}}

\bibitem{Gilmore}
W.~M.~Zhang, D.~H.~Feng  and  R.~Gilmore.  Coherent states: theory and some applications.
{\em Rev.~Mod.~Phys.}  {\bf 62} (1990) 867--927. 
   \href {https://doi.org/10.1103/RevModPhys.62.867}
  {\path{doi:10.1103/RevModPhys.62.867}}

\bibitem{BBF09}
A.~Ballesteros, A.~Blasco A  and F.~J.~Herranz. 
{$N$}-dimensional integrability from two-photon coalgebra symmetry.
{\em J. Phys. A: Math. Theor.} {\bf 42}  (2009) 265205.
    \href {https://doi.org/10.1088/1751-8113/42/26/265205}
  {\path{doi:10.1088/1751-8113/42/26/265205}}  

\bibitem{real}
L.~A.~Real and R.~Biek.  {Spatial dynamics and genetics of infectious diseases on heterogeneous landscapes}. {\it J. Royal Soc. Interface} {\bf 4} (2007) 935--948.  
  \href {https://doi.org/10.1098/rsif.2007.1041}
  {\path{doi:10.1098/rsif.2007.1041}}

\bibitem{duncan}
A.~B.~Duncan, A.~Gonzalez and O.~Kaltz. {Stochastic environmental fluctuations drive epidemiology in experimental host-parasite
metapopulations}. {\it Proc. Royal Soc. B} {\bf 280} (2013) 20131747.  
  \href {https://doi.org/10.1098/rspb.2013.1747}
  {\path{doi:10.1098/rspb.2013.1747}}

\bibitem{kiss}
I.~Z.~Kiss and  P.~L.~Simon. {New moment closures based on a priori distributions with applications to epidemic dynamics}. {\it Bull. Math. Biol.} {\bf 74} (2012) 1501--1515. 
  \href {https://doi.org/10.1007/s11538-012-9723-3}
  {\path{doi:10.1007/s11538-012-9723-3}}
    
\bibitem{Ribe} R.~V.~dos Santos, F.~L.~Ribeiro and A.~S.~Martinez. Models for Allee effect based on physical principles. {\it J. Theor. Biol.} {\bf 385} (2015) 143--152.
\href {https://doi.org/10.1016/j.jtbi.2015.08.018}
 {\path{doi.org/10.1016/j.jtbi.2015.08.018}} 

\bibitem{vilar}
J.~M.~G.~Vilar and  J.~M.~Rubi. Determinants of population responses to environmental fluctuations. {\it Sci. Rep.} {\bf 8} (2018) 887. 
 \href {https://doi.org/10.1038/s41598-017-18976-6}
  {\path{doi:10.1038/s41598-017-18976-6}}
  
\bibitem{Cui}
J.~A.~Cui, X.~Tao and  H.~Zhu.  An SIS infection model incorporating media coverage.  {\it Rocky Mountain J. Math.} {\bf 38} (2008) 1323--1334. 
\href {https://doi.org/10.1216/RMJ-2008-38-5-1323}
  {\path{doi:10.1216/RMJ-2008-38-5-1323}}

\bibitem{Gao}
D.~Gao and  S.~Ruan. An SIS patch model with variable transmission coefficients.  {\it Math. Biosci.} {\bf 232} (2011)  110--115.
\href {https://doi.org/10.1016/j.mbs.2011.05.001}
  {\path{doi:10.1016/j.mbs.2011.05.001}}

\bibitem{Driessche}
P.~van den Driessche. Reproduction numbers of infectious disease models.  {\it Infect. Dis. Mod.} {\bf 2} (2017)  288--303.
\href {https://doi.org/10.1016/j.idm.2017.06.002}
  {\path{doi:10.1016/j.idm.2017.06.002}}
  
\bibitem{LSc} S.~Lie and G.~Scheffers. {\em Vorlesungen {\"u}ber continuierliche {G}ruppen mit geometrischen
  und anderen {A}nwendungen}. (Leipzig: Teubner) 1883.

\bibitem{Car1} J.~F.~Cari\~{n}ena, F.~Falceto and J.~Grabowski. Solvability of a Lie algebra of vector
fields implies their integrability by quadratures.  {\it J. Phys. A: Math. Theor.} {\bf 49} (2016) 425202. 
\href {https://doi.org/10.1088/1751-8113/49/42/425202}
 {\path{doi.org/10.1088/1751-8113/49/42/425202}} 
 
  \bibitem{Car3} J.~F.~Cari\~{n}ena and J.~de Lucas. Lie systems: theory, generalisations, and applications.
{\it Dissertationes Math.} {\bf  479} (2011) 1-162.
\href {https://doi.org/10.4064/dm479-0-1}
 {\path{doi.org/10.4064/dm479-0-1}}   


\bibitem{Car2} J.~F.~Cari\~{n}ena, F.~Falceto, J.~Grabowski and  M.~F.~Ra\~{n}ada. Geometry of Lie integrability
by quadratures.  {\it J. Phys. A: Math. Theor.} {\bf 48} (2015) 215206.
\href {https://doi.org/10.1088/1751-8113/48/21/215206}
 {\path{doi.org/10.1088/1751-8113/48/21/215206}} 
 

\bibitem{Car4}  J.~F.~Cari\~{n}ena, M.~F.~Ra\~{n}ada, F.~Falceto and J.~Grabowski. Revisiting Lie integrability
by quadratures from a geometric perspective. In {\em Geometry of Jets and
Fields}, Banach Center Publ. {\bf 110} (2016) pp.~24--40. Polish Acad. Sci. Inst. Math. Warsaw.
\href {https://doi.org/10.4064/bc110-0-2}
 {\path{doi.org/10.4064/bc110-0-2}}    
  
\bibitem{BC} G. W. Bluman and J. D. Cole. {\em Similarity Methods for Differential Equations}. (New York: Springer) 1974.

\bibitem{Schrod}
A.~Ballesteros,   F.~J.~Herranz and P.~Parashar. 
(1+1) Schr\"odinger Lie bialgebras and their Poisson--Lie groups.
{\em J. Phys. A: Math. Gen.} {\bf 33}  (2000) 3445--3465.
    \href {https://doi.org/10.1088/0305-4470/33/17/304}
  {\path{doi:10.1088/0305-4470/33/17/304}}

 \bibitem{Pra} B.~Prasse and P.~Van Mieghem. Time-dependent solution of the NIMFA equations
around the epidemic threshold. {\it J. Math. Bio.} {\bf 81} (2020) 1299--1355.
\href {https://doi.org/10.1007/s00285-020-01542-6}
 {\path{doi.org/10.1007/s00285-020-01542-6}}    

\bibitem{bon} S.~Bonaccorsi and S.~Ottaviano. A stochastic differential equation SIS
model on network under Markovian switching. {\it Stoch. Anal. Appl.} (2022).
\href {https://doi.org/10.1080/07362994.2022.2146590}
 {\path{doi.org/10.1080/07362994.2022.2146590}}
 
\bibitem{Wi} T.~C.~Bountis, V.~Papageorgiou and P.~Winternitz. On the integrability of systems of nonlinear ordinary differential equations with superposition principles. {\it J. Math. Phys.} {\bf 27} (1986) 1215--1224. 
\href {https://doi.org/10.1063/1.527128}
 {\path{doi.org/10.1063/1.527128}}    
 
\bibitem{Car5} J.~F. Cari\~{n}ena, J.~Grabowski and A.~Ramos. Reduction of time-dependent systems
admitting a superposition principle. {\it  Acta Appl. Math.} {\bf 66} (2001) 67--87.
\href {https://doi.org/10.1023/A:1010743114995}
 {\path{doi.org/10.1023/A:1010743114995}}    

\bibitem{BCFHLb}
A.~Ballesteros, R.~Campoamor-Stursberg, E.~Fern\'andez-Saiz, F.~J.~Herranz and J.~de Lucas.
{Poisson--Hopf algebra deformations of Lie--Hamilton systems.}
 {\it J. Phys. {A}: Math. Theor.} {\bf 51} (2018) 065202.
 \href {https://doi.org/10.1088/1751-8121/aaa090}
  {\path{doi:10.1088/1751-8121/aaa090}}
  
\bibitem{BCFHLc} A.~Ballesteros, R.~Campoamor-Stursberg, E.~Fern\'andez-Saiz, F.~J.~Herranz and  J.~de Lucas. 
A unified approach to Poisson--Hopf deformations of Lie--Hamilton systems based on $\mathfrak{sl}(2)$.
In {\em Quantum theory and symmetries with Lie theory and its applications in physics}, Volume 1, V. Dobrev (ed.), 
Springer Proceedings in Mathematics \& Statistics {\bf 263} (2018), pp.~347--366.
  \newblock \href {https://doi.org/10.1007/978-981-13-2715-5_23}
  {\path{doi:10.1007/978-981-13-2715-5_23}}

\bibitem{CP}   V.~Chari and A.~Pressley. {\em A Guide to Quantum Groups}. (Cambridge: Cambridge University Press) 1994.

\bibitem{h4}  A.~Ballesteros and  F.~J.~Herranz. Lie bialgebra quantizations of the oscillator algebra and their universal $R$-matrices.
 {\it J. Phys. A: Math. Gen.} {\bf 29} (1996)  4307--4320. 
   \href {https://doi.org/10.1088/0305-4470/29/15/006}
  {\path{doi:10.1088/0305-4470/29/15/006}}

    
\end{thebibliography}
\end{document}